\documentclass[submission,copyright,creativecommons]{eptcs}
\providecommand{\event}{HSB 2012}
\usepackage{fixme}
\usepackage[ruled,vlined]{algorithm2e}
\usepackage[utf8]{inputenc}
\usepackage{amssymb,
            amsmath,amsthm,
            mathrsfs,
            url,
            color,
            ifpdf,
            setspace,
            semantic,
            xspace,
            pstricks,
            pst-node,
            booktabs,
            wrapfig,
            colortbl}
\usepackage[final]{graphicx}
\usepackage{hyperref}

\usepackage{wrapfig}
\usepackage{caption,subcaption} 
\graphicspath{{fig/}}

\usepackage{amssymb}
\usepackage{xspace}
\usepackage{textcomp} 
\usepackage{calc}

\DeclareMathAlphabet\mathscra{T1}{hlcw}{m}{it}

\let\then\iftrue
{\catcode`\!=8 
 \catcode`\Q=3 
 \catcode`\@=11
 \long\gdef\ifgiven#1\then{\Ifbl@nk#1QQQ\empty!}
 \long\gdef\ifblank#1\then{\Ifbl@nk#1QQ..!}
 \long\gdef\Ifbl@nk#1#2Q#3!{\ifx#3}
 \long\gdef\ifnull#1\then{\IfN@LL#1* {#1}!}
 \long\gdef\IfN@LL#1 #2!{\ifblank{#2}\then}}

    {\bigskip\noindent\begin{tabular}{|p{\textwidth}}\it\small\strut}%
    {\strut\end{tabular}\par\bigskip}

    {\begingroup\vskip-2ex\begin{small}\noindent}%
    {\end{small}\vskip1.5mm\endgroup}

    {\begingroup\small\begin{equation}}%
    {\end{equation}\endgroup}
\newcommand*{\stignore}[1]{}

\newcommand{\wordsp}[1]{\textit{#1}}

\newcommand*{\mktype}[2]{%
  \expandafter\newcommand\csname #1sym\endcsname%
      {{\textsf{#2}}}%
  \expandafter\newcommand\csname #1name\endcsname%
      {{\csname #1sym\endcsname}\xspace}%
  \expandafter\newcommand\csname #1\endcsname%
      {{\csname #1sym\endcsname}\xspace}%
    }

\newcommand*{\mkkeyword}[2]{%
  \expandafter\newcommand\csname #1sym\endcsname%
      {{\ensuremath{#2}\expandafter\index{#2@\protect{\csname #1sym\endcsname}}}}%
  \expandafter\newcommand\csname #1name\endcsname%
      {{\csname #1sym\endcsname}\xspace}%
  \expandafter\newcommand\csname #1\endcsname%
      {{\csname #1sym\endcsname}\xspace}%
    }

\newcommand*{\mkfunc}[3]{%
  \expandafter\newcommand\csname #1sym\endcsname%
  {\ensuremath{#2}\index{#1@\protect{\csname #1sym\endcsname}}}%
  \expandafter\newcommand\csname #1name\endcsname%
  {\csname #1sym\endcsname\xspace}%
  \expandafter\newcommand\csname #1\endcsname[1]%
  {\ensuremath{\csname #1sym\endcsname #3{##1}}\xspace}%
}

\newcommand{\encloseinpar}[1]{(#1)}

\newcommand{\encloseinsembra}[1]{|[#1|]}
\newcommand{\encloseinangles}[1]{\langle #1\rangle}

\newcommand{\mksubscript}[1]{_{#1}}
\newcommand{\donotenclose}[1]{#1}
\newcommand{\followwithbang}[1]{#1{\ensuremath{\scriptstyle !}}}
\newcommand{\followwithqstn}[1]{#1{\ensuremath{\scriptstyle ?}}}
\newcommand{\followwithnot}[1]{#1{\raisebox{-0.8ex}[0mm][.1ex]{\ensuremath{\scriptstyle \!\not\,\,}}}}
\newcommand{\followwithast}[1]{#1^\ast}

\newcommand*{\mkfuncn}[2]{\mkfunc{#1}{#2}{\donotenclose}}    
\newcommand*{\mkfuncp}[2]{\mkfunc{#1}{#2}{\encloseinpar}}    
\newcommand*{\mkfuncs}[2]{\mkfunc{#1}{#2}{\encloseinsembra}} 
\newcommand*{\mkfuncsub}[2]{\mkfunc{#1}{#2}{\mksubscript}}   

\newcommand*{\mkbinaryrel}[2]{%
  \expandafter\newcommand\csname#1sym\endcsname%
  {{\ensuremath{#2}}}%
  \expandafter\newcommand\csname not#1sym\endcsname%
  {{\ensuremath{\not{#2}}}}%
  \expandafter\newcommand\csname #1name\endcsname%
  {\csname #1sym\endcsname\xspace}%
  \expandafter\newcommand\csname not#1name\endcsname%
  {\csname not#1sym\endcsname\xspace}%
  \expandafter\newcommand\csname #1\endcsname[2]%
  {{\ensuremath{##1#2##2}}}%
  \expandafter\newcommand\csname not#1\endcsname[2]%
  {{\ensuremath{##1 \not#2  ##2}}}%
}

\newcommand*{\mkternaryrel}[2]{%
  \expandafter\newcommand\csname #1sym\endcsname%
  {\ensuremath{#2}}%
  \expandafter\newcommand\csname not#1sym\endcsname%
  {\ensuremath{\not #2}}%
  \expandafter\newcommand\csname #1name\endcsname%
  {\csname #1sym\endcsname\xspace}%
  \expandafter\newcommand\csname not#1name\endcsname%
  {\csname not#1sym\endcsname\xspace}%
  \expandafter\newcommand\csname #1\endcsname[3]%
  {\ensuremath{##1 \csname #1sym\endcsname_{##2} ##3}}%
  \expandafter\newcommand\csname not#1\endcsname[3]%
  {\ensuremath{##1 \not\csname #1sym\endcsname_{##2} ##3}}%
}

\newcommand*{\mkternaryrelraised}[2]{%
  \expandafter\newcommand\csname #1sym\endcsname%
  {\ensuremath{#2}}%
  \expandafter\newcommand\csname not#1sym\endcsname%
  {\ensuremath{\not #2}}%
  \expandafter\newcommand\csname #1name\endcsname%
  {\csname #1sym\endcsname\xspace}%
  \expandafter\newcommand\csname not#1name\endcsname%
  {\csname not#1sym\endcsname\xspace}%
  \expandafter\newcommand\csname #1\endcsname[3]%
  {\ensuremath{##1 \csname #1sym\endcsname{}^{##2} ##3}}%
  \expandafter\newcommand\csname not#1\endcsname[3]%
  {\ensuremath{##1 \not\csname #1sym\endcsname{}^{##2} ##3}}%
}

\newcommand*{\mkcomrel}[5]{%
  \ifgiven{#4}\then%
    \expandafter\newcommand\csname #1sym\endcsname%
    {\raisebox{.2ex}{\ensuremath{\xrightarrow[{\raisebox{.5ex}[0mm][.1ex]{\tiny #4}}]%
          {#3{}}}}{}%
    \index{#4@\protect{\csname #1sym\endcsname}}}%
    \expandafter\newcommand\csname #1name\endcsname%
    {\csname #1sym\endcsname\xspace}%
    \expandafter\newcommand\csname #1\endcsname[3]%
    {\hbox{\ensuremath{%
          #2{##1}%
          \raisebox{-.1ex}{\ensuremath{\xrightarrow[{\raisebox{.5ex}[0mm][.1ex]{\smash{\tiny #4}}}]%
              {\phantom{.}\raisebox{-0.3ex}[0mm][0.1ex]{{#3{\ensuremath{\scriptstyle ##2}}}}\phantom{.}}}}%
          {}#5{##3}%
        }%
      }}%
  \else%
    \expandafter\newcommand\csname #1sym\endcsname%
    {\ensuremath{\xrightarrow{#3{}}}%
      \index{#4@\protect{\csname #1sym\endcsname}}}%
    \expandafter\newcommand\csname #1name\endcsname%
    {\csname #1sym\endcsname\xspace}%
    \expandafter\newcommand\csname #1\endcsname[3]%
    {\hbox{\ensuremath{%
          #2{##1}%
          \ensuremath{\xrightarrow{#3{##2}}}%
          #5{##3}%
        }%
      }\index{#4@\protect{\csname #1sym\endcsname}}}%
  \fi%
}

\newcommand*{\mkoperel}[4]{%
  \expandafter\newcommand\csname #1sym\endcsname%
       {\raisebox{.2ex}{\ensuremath{\xrightarrow[{\raisebox{.5ex}[0mm][.1ex]{\tiny #3}}]%
             {}}}\index{#3@\protect{\csname #1sym\endcsname}}}%
  \expandafter\newcommand\csname #1name\endcsname%
       {\csname #1sym\endcsname\xspace}%
  \expandafter\newcommand\csname #1\endcsname[2]%
       {\hbox{\ensuremath{%
              #2{##1}%
              \raisebox{.2ex}{\ensuremath{\xrightarrow[{\raisebox{.5ex}[0mm][.1ex]{\tiny #3}}]{}}}%
              #4{##2}%
            }%
          }\index{#3@\protect{\csname #1sym\endcsname}}}%
}

\newcommand*{\mkoperelaa}[2]{%
  \mkoperel{#1}{\encloseinangles}{#2}{\encloseinangles}}                      
\newcommand*{\mkcomrelaa}[2]{%
  \mkcomrel{#1}{\encloseinangles}{\donotenclose}{#2}{\encloseinangles}}   
\newcommand*{\mkoperelnn}[2]{%
  \mkoperel{#1}{\donotenclose}{#2}{\donotenclose}}                            
\newcommand*{\mkcomrelnn}[2]{%
  \mkcomrel{#1}{\donotenclose}{\donotenclose}{#2}{\donotenclose}}         
\newcommand*{\mkoperelan}[2]{%
  \mkoperel{#1}{\encloseinangles}{#2}{\donotenclose}}                         
\newcommand*{\mkcomrelan}[2]{%
  \mkcomrel{#1}{\encloseinangles}{\donotenclose}{#2}{\donotenclose}}      

\newcommand*{\mkdirrelaa}[3]{%
  \mkcomrel{#1}{\donotenclose}{#2}{#3}{\donotenclose}}         

\newcommand{\VSsym}{\index{visualSTATE@\protect{\VSsym}}\textsf{visual\-STATE}}

\newcommand{\STM}{\index{Statemate@\protect{\STM}}\textsc{statemate}\xspace}
\newcommand{\SCsym}{\expandafter\index{SCOPE@\protect{\SCsym}}\textsf{SCOPE}}

\newcommand{\Rhapsodysym}{\index{Rhapsody@\protect{\Rhapsodysym}}\textsc{Rhapsody}}

\newcommand{\inonek}{\hbox{\ensuremath{\in\kern-.5mm\{1..k\}}}}

\mkfuncp{elems}{\wordsp{elems}}
\mkfuncp{dom}{\wordsp{dom}}
\mkfuncp{rng}{\wordsp{rng}}
\mkfuncp{Power}{\mathscr{P}}
\mkfuncp{Multi}{\mathcal{M}}
\mkfuncp{Forest}{\mathcal{F}}

\mkfunc{List}{}{\followwithast}
\newcommand{\Projectsym}[1]{\ensuremath{\pi_{#1}}}
\newcommand{\Project}[2]
        {\ensuremath{\Projectsym{#1}({#2})}\xspace}

\mktype{OR}{or}             
\mktype{XOR}{xor}           
\mktype{AND}{and}    

\mkkeyword{Ids}{\wordsp{Id}}
\mkkeyword{State}{\wordsp{State}}
\mkkeyword{History}{\wordsp{History}}

\newcommand*{\ORst}{\expandafter\index{or-state@\protect{\ORst}}\ORsym{}-state\xspace}
\newcommand*{\ORsts}{\expandafter\index{or-state@\protect{\ORst}}\ORsym{}-states\xspace}
\newcommand*{\XORst}{\expandafter\index{xor-state@\protect{\XORst}}\XORsym{}-state\xspace}
\newcommand*{\XORsts}{\expandafter\index{xor-state@\protect{\XORst}}\XORsym{}-states\xspace}
\newcommand*{\ANDst}{\expandafter\index{and-state@\protect{\ANDst}}\ANDsym{}-state\xspace}
\newcommand*{\ANDsts}{\expandafter\index{and-state@\protect{\ANDst}}\ANDsym{}-states\xspace}

\newcommand*{\StateORsym}{\index{\protect{\StateORsym}}\ensuremath{\State_{\text{\ORsym}}}}   
\newcommand*{\StateANDsym}{\index{\protect{\StateANDsym}}\ensuremath{\State_{\text{\ANDsym}}}}

\mkfuncp{type}{\Gamma_S}

\newcommand{\sbstatekern}{\kern -0.18em}

\newcommand{\sbstatesym}{\ensuremath{\searrow}\xspace}  
\newcommand{\notsbstatesym}{\ensuremath{\not\searrow}\xspace}

\newcommand{\sbstateplsym}{\ensuremath{\sbstatesym\kern -0.66em^{+}\kern 0.13em}}
 
\newcommand{\notsbstatepl}[2]{\hbox{\ensuremath{#2\sbstatekern\notsbstatesym\kern -0.66em^{+}#1}}\xspace} 
\newcommand{\sbstatestsym}{\ensuremath{\sbstatesym\kern -0.66em^\ast\kern 0.13em}}
 
\newcommand{\notsbstatest}[2]{\hbox{\ensuremath{#2\sbstatekern\notsbstatesym\kern -0.66em^\ast#1}}\xspace}

\mkfuncp{priority}{\vartriangleleft}
\mkfuncp{stpriority}{\blacktriangleleft}

\mkfuncp{parent}{\wordsp{parent}}
\mkfuncp{parparent}{\wordsp{parent}^2}
\mkfuncp{children}{\wordsp{children}}
\mkfuncp{ancestors}{\wordsp{ancest}}
\mkfuncp{descendants}{\wordsp{descend}}
\mkfuncp{siblings}{\wordsp{siblings}}

\mkfuncp{ancestorspl}{\wordsp\ancestorssym^{+}}
\mkfuncp{descendantspl}{\wordsp\descendantssym^{+}}
\mkfuncp{ancestorsst}{\wordsp\ancestorssym^\ast}
\mkfuncp{descendantsst}{\wordsp\descendantssym^\ast}
\mkfuncp{siblingsst}{\wordsp{\siblingssym}^\ast}

\mkfuncp{NCA}{\wordsp{NCA}}

\newcommand{\notorthogonalsym}{\hbox{\ensuremath{\not{\kern -2.2pt\bot\kern +2.2pt}}}}

\mkfuncp{initial}{\wordsp{ini}}
\mkfuncp{default}{\wordsp{default}}
\mkfuncp{runhistory}{\eta}

\mkfuncp{entrypathOR}{\entrypathsym_{\text{\ORsym}}}
\mkfuncp{entrypathAND}{\entrypathsym_{\text{\ANDsym}}}
\mkfuncp{entrypath}{\wordsp{entr}}
\mkfuncp{entrysigs}{\wordsp{ensigs}}

\mkfuncp{entrORts}{\entrypathORsym\kern-6pt^{\wordsp{ts}}}
\mkfuncp{entrANDts}{\entrypathANDsym\kern-10pt^{\wordsp{ts~}}}
\mkfuncp{exitpath}{\wordsp{extr}}
\mkfuncp{exitsigs}{\wordsp{exsigs}}

\mkkeyword{Id}{\wordsp{Id}}
\mkkeyword{Var}{\wordsp{Var}}
\mkkeyword{Value}{\wordsp{Value}}

\mkkeyword{ExtVar}{\Var_E}
\mkkeyword{IntVar}{\Var_I}
\mkkeyword{Type}{\wordsp{Type}}
\mkkeyword{SimpleType}{\wordsp{SimpleType}}

\mkkeyword{anatomy}{\wordsp{anatomy}}

\mktype{tint}{int}
\mktype{tuint}{uint}
\mktype{treal}{real}
\mktype{tfloat}{float}
\mktype{tdouble}{double}
\mktype{tvoid}{void}

\mkfuncp{Dom}{D}

\mkfuncp{vartype}{\Gamma_V}
\mkfuncp{Array}{\wordsp{Vector}}
\mkfuncs{toracle}{\tau}

\mkfuncp{mustclosure}{\raisebox{-2pt}{$\Box$}\textit{-closure}}

\mkkeyword{Event}{\wordsp{Event}}
\mkkeyword{Signal}{\wordsp{Signal}} 
\mkkeyword{External}{\wordsp{External}} 
\mkkeyword{Guard}{\wordsp{Guard}}
\mkkeyword{Ebind}{\wordsp{Ebind}}
\mkkeyword{Einst}{\wordsp{Einst}}
\mkkeyword{Sexp}{\wordsp{Sexpr}}
\mkkeyword{Lvalue}{\wordsp{Access}}

\mkfuncp{evtype}{\Gamma_E}

\mkkeyword{topstate}{\wordsp{root}}

\mkkeyword{Exp}{\wordsp{Exp}}
\mkkeyword{OExp}{\wordsp{exp}}
\mkkeyword{Aexp}{\wordsp{Aexp}}
\mkkeyword{Bexp}{\wordsp{Bexp}}

\mkkeyword{Assgn}{\wordsp{Assgn}}
\mkkeyword{Output}{\wordsp{Output}}
\mkkeyword{Oinst}{\wordsp{Oinst}}
\mkkeyword{Ainst}{\wordsp{Ainst}}
\mkkeyword{Oexp}{\wordsp{Oexpr}}

\mkfuncp{fntype}{\Gamma_F}

\mkkeyword{Action}{\wordsp{Action}}
\mkkeyword{Trans}{\wordsp{Trans}}

\mkfuncp{params}{\wordsp{params}}
\mkfuncp{scope}{\wordsp{scope}}
\mkfuncp{iscope}{\wordsp{iscope}}
\mkfuncp{scopes}{\wordsp{scope}}
\mkfuncp{poscond}{\wordsp{pos}}
\mkfuncp{negcond}{\wordsp{neg}}
\mkfuncp{targets}{\wordsp{targets}}
\mkfuncp{source}{\wordsp{source}}
\mkfuncp{expr}{\wordsp{expr}}
\mkfuncp{guard}{\wordsp{guard}}
\mkfuncp{entry}{\wordsp{en}}
\mkfuncp{exit}{\wordsp{ex}}

\mkfuncp{freevars}{\wordsp{fvars}}
\mkfuncp{pure}{\wordsp{pure}}
\mkfuncp{closed}{\wordsp{closed}}

\mkkeyword{Store}{\wordsp{Store}}
\mkkeyword{Queue}{\wordsp{Queue}}
\mkkeyword{state}{\sigma}
\mkkeyword{gstate}{\Sigma}

\mkfuncp{store}{\varrho}
\mkfuncp{history}{\wordsp{his}}

\mkkeyword{queue}{q}

\mkoperelnn{trarrow}{\stignore{0transitionrelation}}
\mkoperelaa{access}{\textrm{access}}
\mkoperelaa{Aeval}{\textrm{Aeval}}
\mkoperelan{Aexec}{\textrm{asgn}}
\mkoperelan{Beval}{\textrm{Beval}}
\mkoperelaa{Cexec}{\textrm{C-sem}}
\mkcomrelnn{comarrow}{}
\mkcomrelnn{Init}{\textrm{init}}
\mkcomrelaa{Macro}{\textrm{macro}}
\mkcomrelaa{Micro}{\textrm{micro}}
\mkoperelaa{Mini}{\textrm{mini}}
\mkcomrelaa{Oeval}{\textrm{output}}
\mkcomrelan{Rexec}{\textrm{exec}}  
\mkcomrelnn{Senter}{\textrm{enter}}
\mkoperelaa{Seval}{\textrm{signal}}
\mkcomrelnn{Sexit}{\textrm{exit}}
\mkcomrelaa{systr}{\stignore{0transitionrelation}}
\mkcomrelnn{Tfire}{\textrm{fire}}

\mkfuncp{cons}{\wordsp{cons}}

\mkfuncn{tail}{\text{\bfseries tl}\,}

\mkfuncn{etail}{{\textbf{src}}\,}
\mkfuncn{ehead}{{\textbf{tgt}}\,}

\mkfuncp{Br}{\textsf{Br}}

\mkfuncp{activeset}{\wordsp{enabled}}
\mkfuncp{enabled}{\wordsp{enabled}}
\mkfuncp{resactive}{\wordsp{resolved-enabled}}
\mkfuncp{activestate}{\wordsp{active}}
\mkfuncp{inactivestate}{\wordsp{inactive}}

\mkfuncp{traces}{\wordsp{traces}}

\mkkeyword{PR}{\wordsp{Pr}}
\mkkeyword{PRP}{\PRsym_{\mathcal{P}}}
\mkkeyword{PRS}{\PRsym_{\ast}}
\mkkeyword{ENV}{\wordsp{Env}}
\mkkeyword{GEN}{\wordsp{Gen}}
\mkkeyword{GENP}{\GENsym_\mathcal{P}}
\mkkeyword{GENS}{\GENsym_\ast}
\mkkeyword{OBS}{\wordsp{Obs}}
\mkkeyword{OBSP}{\OBSsym_\mathcal{P}}
\mkkeyword{OBSS}{{\OBSsym_\ast}}

\mkkeyword{OUT}{\wordsp{Out}}
\mkkeyword{IN}{\wordsp{In}}

\newcommand{\gentrsymname}{}
\newcommand{\obstrsymname}{}
\newcommand{\nottrsymname}{}
\mkdirrelaa{gentr}{\followwithbang}{\gentrsymname}
\mkdirrelaa{obstr}{\followwithqstn}{\obstrsymname}
\mkdirrelaa{nottr}{\followwithnot}{\nottrsymname}

\mkternaryrel{relsim}{\leqslant}
\mkternaryrel{relbisim}{\sim}
\mkternaryrel{twowayrelsim}{\index{simulation!relativized, two way}\lessgtr}
\mkbinaryrel{twowaysim}{\index{simulation!two way}\lessgtr}
\mkbinaryrel{osim}{\leqslant}
\mkbinaryrel{gsim}{\leqslant}
\mkbinaryrel{obisim}{\sim}
\mkbinaryrel{gbisim}{\sim}
\mkbinaryrel{syssim}{\leqslant}
\mkbinaryrel{modref}{\leq_{\textnormal{\textrm{m}}}}
\mkbinaryrel{altsim}{\leq_{\textnormal{\textrm{a}}}}
\mkternaryrelraised{omodref}{\leq^{\ast}_{\textnormal{\textrm{m}}}}
\mkternaryrelraised{agmodref}{\leq^{\ast}_{\textnormal{\textrm{ag}}}}
\mkternaryrelraised{esopref}{\leq^{\triangleleft}_{\textnormal{\textrm{m}}}}
\mkbinaryrel{oaltsim}{\leq^{\ast}_{\textnormal{\textrm{a}}}}
\mkbinaryrel{envsim}{\leqslant}
\mkbinaryrel{sysbisim}{\sim}
\mkbinaryrel{discr}{\index{discrimination}\sqsubseteq}
\mkbinaryrel{implements}{\preceq}
\mkbinaryrel{twodiscr}{\raisebox{1.25ex}{\rotatebox[origin=ct]{-180}%
    {\index{discrimination}\smash{\ensuremath{\sqsupseteq}}}}}
\mkbinaryrel{bidiscr}{\raisebox{-.5ex}{\ensuremath{%
      \index{discrimination}\stackrel{\sqsubset}{\scalebox{0.8}{\ensuremath{\sim}}}}}}

\mkbinaryrel{composetr}{\!\upharpoonright\!}
\mkbinaryrel{crossprod}{\otimes}

\mkkeyword{IOATS}{\text{\index{IOATS}IOATS}}
\mkkeyword{IOATSs}{\text{\index{IOATS}IOATSs}}

\mkfuncp{SSF}{\mathbb{S}}

\mkkeyword{SET}{\textbf{Set}}
\mkkeyword{set}{\textbf{set}}
\mkkeyword{MSET}{\textbf{M-set}}
\mkkeyword{mset}{\textbf{m-set}}
\mkkeyword{INTER}{\textnormal{\textbf{Inter}}}
\mkkeyword{inter}{\textnormal{\textbf{inter}}}

\newcommand{\BLINDsym}{\ensuremath{\index{environment!blind}\mathbf{B}}}

\mkfuncsub{parBLIND}{\BLINDsym}

\newcommand{\UNIVsym}{{\ensuremath{\index{environment!perfect vision}\mathbf{V}}}}

\mkfuncsub{parUNIV}{\UNIVsym}

\mkfuncp{eqdom}{\wordsp{part}}

\usepackage{psfrag}
\usepackage{ifthen}

\newtheorem{definition}{Definition}
\newtheorem{example}{Example}

\newcommand{\uppaal}{{\sc Uppaal}\xspace}
\newcommand{\uppaalsmc}{{\sc Uppaal-smc}\xspace}

\newcommand{\arrow}[1]{\stackrel{#1}{\longrightarrow}}

\newcommand{\St}{\textit{St}\xspace}

\renewcommand{\P}{\ensuremath{{\mathbb P}}\xspace}

\newcommand{\xRightarrow}[2][]{%
     \ext@arrow 0055{\Rightarrowfill@}{#1}{#2}%
} 

\renewcommand{\St}{{\sf St}}
\newcommand{\A}{\boldsymbol{{\cal A}}}

\renewcommand{\phi}{\varphi}

\newcommand{\forget}[1]{}

\newcommand{\bbbr}{\ensuremath{{\mathbb R}}\xspace}

\newcommand{\rplus}{\ensuremath{\bbbr_{\geq0}}}

\def\titlerunning{Statistical  Model  Checking  for Stochastic  Hybrid  Systems}
\def\authorrunning{David, Du, Larsen, Legay, Miku\v{c}ionis, Poulsen, Sedwards}

\begin{document}

\title{Statistical  Model  Checking  for Stochastic  Hybrid  Systems
\thanks{Work  partially supported by  the VKR Centre  of Excellence
 MT-LAB, the Sino-Danish Basic Research Center IDEA4CPS, and by the
 CREATIVE project ESTASE.}
}

\author{
Alexandre~David\quad
Kim~G.~Larsen\quad
Marius~Miku\v{c}ionis\quad
Danny~B{\o}gsted~Poulsen
\institute{Department of Computer Science\\
Aalborg University, Denmark}
\email{\{adavid,kgl,marius,dannybp\}@cs.aau.dk}
\and
Axel~Legay\quad
Sean~Sedwards
\institute{INRIA Rennes\\
Bretagne Atlantique, France}
\email{axel.legay@inria.fr,sean.sedwards@gmail.com}
\and
Dehui~Du
\institute{Shanghai Key Laboratory of Trustworthy Computing\\
East China Normal University, \\
Shanghai {\rm 20062}, China}
\email{dehuidu@cs.aau.dk}
}

\maketitle

\begin{abstract}
  This  paper  presents  novel  extensions  and  applications  of  the
  \uppaalsmc model checker. The extensions allow for statistical model
  checking of  stochastic hybrid systems.  We show  how our race-based
  stochastic  semantics extends  to  networks of  hybrid systems,  and
  indicate  the integration  technique applied  for  implementing this
  semantics  in the \uppaalsmc  simulation engine.   We report  on two
  applications of  the resulting tool-set coming  from systems biology
  and energy aware buildings.
\end{abstract}

\section{Introduction}

Statistical                       Model                       Checking
(SMC)\,\cite{LDB10,LMPR07,SVA04,YS02,KZHHJ11} is  an approach that has
recently been  proposed as  new validation technique  for large-scale,
complex systems. The  core idea of SMC is  to conduct some simulations
of  the  system,  monitor  them,  and  then  use  statistical  methods
(including sequential hypothesis testing or Monte Carlo simulation) in
order  to decide  with some  degree of  confidence whether  the system
satisfies the property or not.  By nature, SMC is a compromise between
testing  and  classical  formal method  techniques.   Simulation-based
methods  are known  to  be far  less  memory and  time intensive  than
exhaustive ones,  and are  some times the  only option.  SMC  has been
implemented in a  series of tools that have  defeated well-known tools
such  as  PRISM  on  several  case studies.   Unlike  more  “academic”
exhaustive (and intractable) techniques,  SMC is spreading to various research
areas  such as  systems  biology\,\cite{GZKFC10,JCLLPZ09} and  software
engineering\,\cite{ZPC10,MPL11},   in    particular   for   industrial
applications\,\cite{BBBCDL10,BBBDLS10,DBLP:journals/sttt/BasuBBDL12}.

There are several reasons for this success.  Firstly, SMC is simple to
understand,  use  and  (in  principle)  to  implement.   Secondly,  no
additional modelling  or specification effort is  needed, provided that
the  mode-ling  formalism  used  can  be  given  a  natural  stochastic
semantics serving as the basis for interpretation of the specification
formalism and as a basis for generating simulation-runs.  Thirdly, SMC
allows  to analyse  properties\,\cite{CDL10,BBBCDL10}  that cannot  be
expressed in classical temporal  logic, including properties for which
classical model checking is undecidable.

In  a  series   of  recent  works\,\cite{DLLMW11,DLLMPVW11},  we  have
investigated the problem of Statistical Model Checking for networks of
Priced Timed  Automata (PTAs).  PTAs are timed  automata, whose clocks
can  evolve with  different rates,  while\footnote{in contrast  to the
  usual restriction of priced timed automata \cite{PTA01,WTA01}} being
used   with   no   restrictions   in  guards   and   invariants.    In
\cite{DLLMPVW11}, we have proposed  a natural stochastic semantics for
such automata, which allows to perform statistical model checking. Our
work has  later been implemented  in \uppaalsmc, that is  a stochastic
and statistical model checking extension of \uppaal. \uppaalsmc relies
on a series  of extensions of the statistical  model checking approach
generalised  to  handle  real-time  systems and  estimate  undecidable
problems.  \uppaalsmc  comes together  with a friendly  user interface
that allows a user to  specify complex problems in an efficient manner
as well  as to get feedback  in the form  of probability distributions
and compare  probabilities to analyse performance  aspects of systems.
The  \uppaalsmc model checking  has been  applied to  a wide  range of
examples from networking  and  Nash equilibrium\cite{BDLLM12}  through
systems                biology\,\cite{ourbio},               real-time
scheduling\,\cite{ourherschel},         and        energy        aware
systems\,\cite{chinesepaper}.

For PTAs clocks evolve with fixed rates depending only on the discrete
state of the  model, i.e.  locations and discrete  variables.  In this
paper, we present and implement an extension of the modelling formalism
of  \uppaalsmc, where clock  rates may  depend not  only on  values of
discrete variables but also on  the value of other clocks, effectively
amounting  to  ordinary differential  equations  (ODEs).   As a  first
contribution of this paper, we  present an extension of our race-based
stochastic semantics to networks of stochastic hybrid automata (SHA).
%
%
One major difficulty  is the implementation of this  extensions in our
engine for generating  random runs, in a manner  which is correct with
respect  to the  stochastic semantics.   In fact  \uppaalsmc  does not
solve  the equations  exactly  but   currently    supports  the  Euler
integration method.  A  fixed time step (defined by  the user) is used
by  an  internal  integrator  component  added to  the  system.   This
integrator races with the other processes during its time step and all
rates  are considered  constant and  defined by  the equations  in the
model.   As  a second  contribution  of  the  paper we  describe  this
integration in  \uppaalsmc.  It is  worth mentioning that  while other
SMC-based approaches  exist for SHAs\,\cite{ZPC10,ZBC12,BDDHP11}, none
of them do consider ODE-based modelisation for clock rates.

To  demonstrate the  applicability  of this  new  extension, we  apply
advanced SMC  techniques to  two challenging applications  coming from
systems biology  and energy aware  buildings. For the case  of systems
biology, we  show how  the combination  of ODEs and  SMC allows  us to
reason  on biological  oscillations --  a problem  that is  beyond the
scope  of most existing  formal verification  techniques.  We  model a
genetic circadian oscillator, which is  used to distil the essence of
several  real circadian  oscillators.  For  the case  of  energy aware
buildings,  we  refer  to  a recently  developed  framework  including
components for  layout of buildings, availability  of heaters, climate
and  user behaviours  allowing  to evaluate  different strategies  for
distributing heaters  among rooms in  terms of the  resulting comfort
and energy  consumption.  To indicate central parts  of this framework
and  the   clear  advantages  of  modelling  the   evaluation  of  room
temperatures with ODEs, we illustrate in this paper the framework with
a small instance comprising two rooms with a single shared heater.

\paragraph{Structure of the Paper.} 
The  remainder of the  paper is  structured as  follows.  In  the next
Section~\ref{sec:ball} we  preview the expressive power  of the hybrid
extensions of \uppaalsmc using an extension of the well-known bouncing
ball.  Sections~\ref{sec:nha} and  \ref{sec:sha} details the semantics
of the extended formalism  of networks of stochastic hybrid automata.
Applications to Energy Aware Buildings and a Biological Oscillator are
given   in   Section   \ref{build-sec}  and   Section   \ref{bio-sec},
respectively. Finally,  Section~\ref{con-sec} concludes the  paper and
suggests directions for future research.

\section{Throwing, Bouncing and Hitting Ball}
\label{sec:ball}
To give an early illustration  of the expressive power of the extended
modelling  formalism  of  \uppaalsmc,  we  consider a  variant  of  the
well-known  bouncing  ball.  
\begin{figure}[tb]
  \begin{subfigure}[b]{0.34\linewidth}
    \includegraphics[trim=0 0 4em 0 0 cut,width=\linewidth]{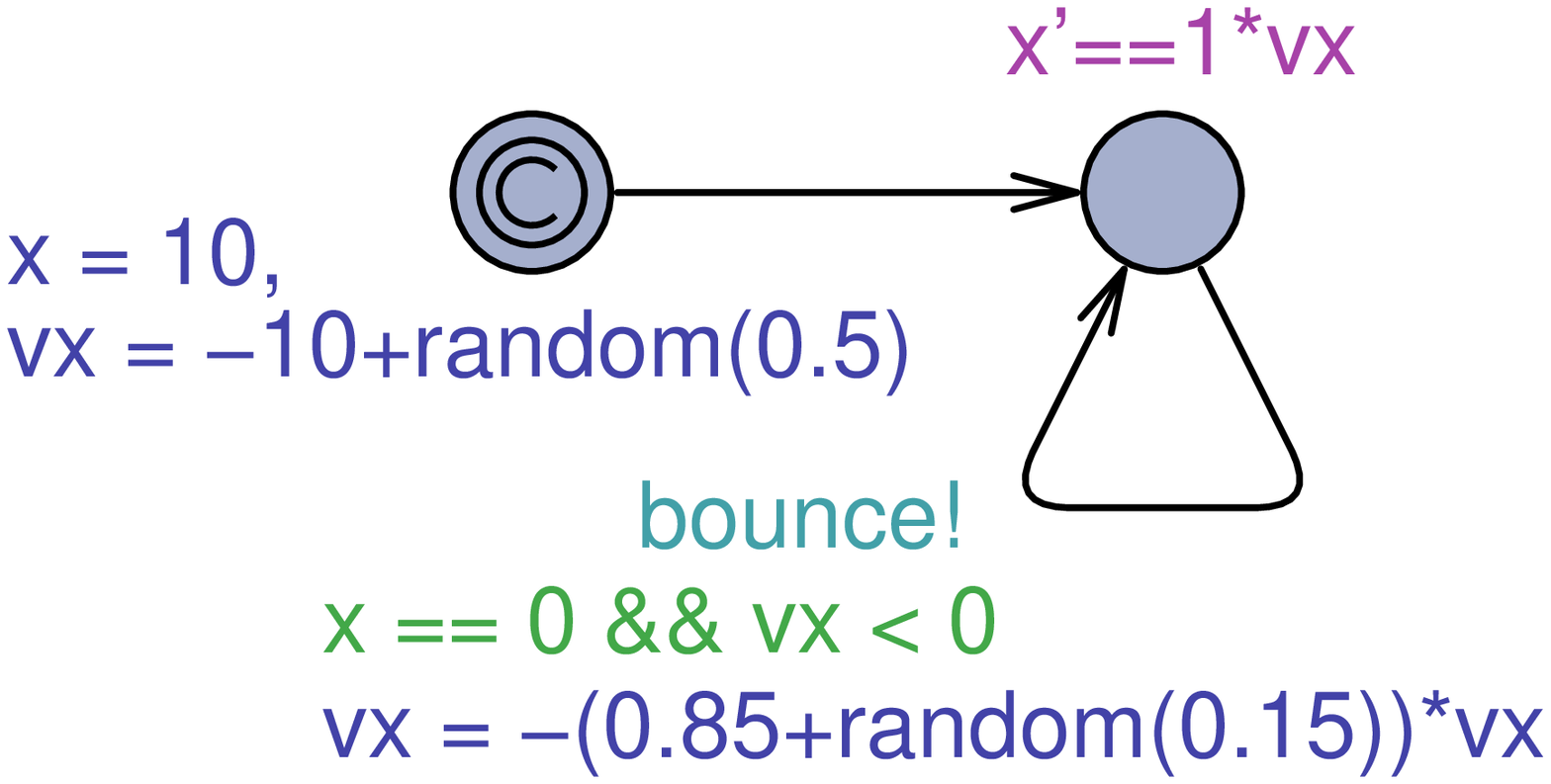}
    \caption{Model of $x$ coordinate.}
    \label{fig:ballx}
  \end{subfigure}\,\,\,
  \begin{subfigure}[b]{0.60\linewidth}  
    \includegraphics[trim=0 0 3em 0 0 cut,width=\linewidth]{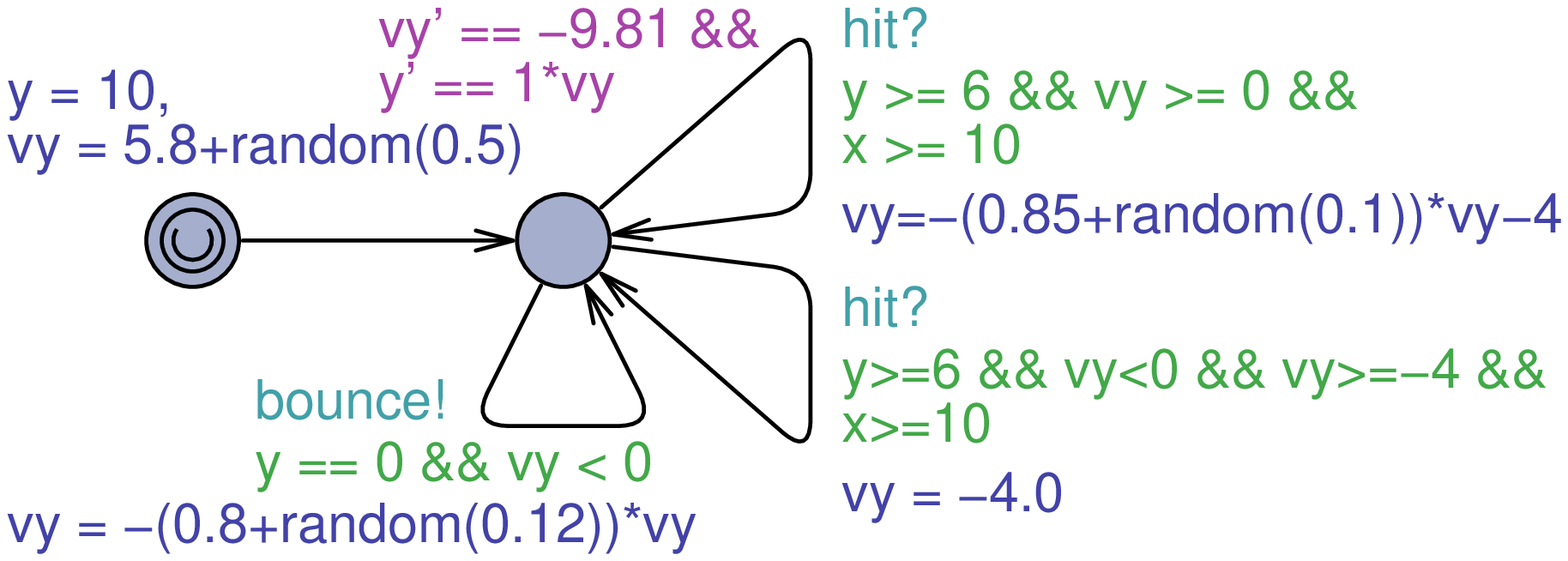} 
    \caption{Model of $y$ coordinate.}
    \label{fig:bally}
  \end{subfigure}\\
  \begin{subfigure}[b]{0.37\linewidth}
    \centering
    \includegraphics[width=0.286\linewidth]{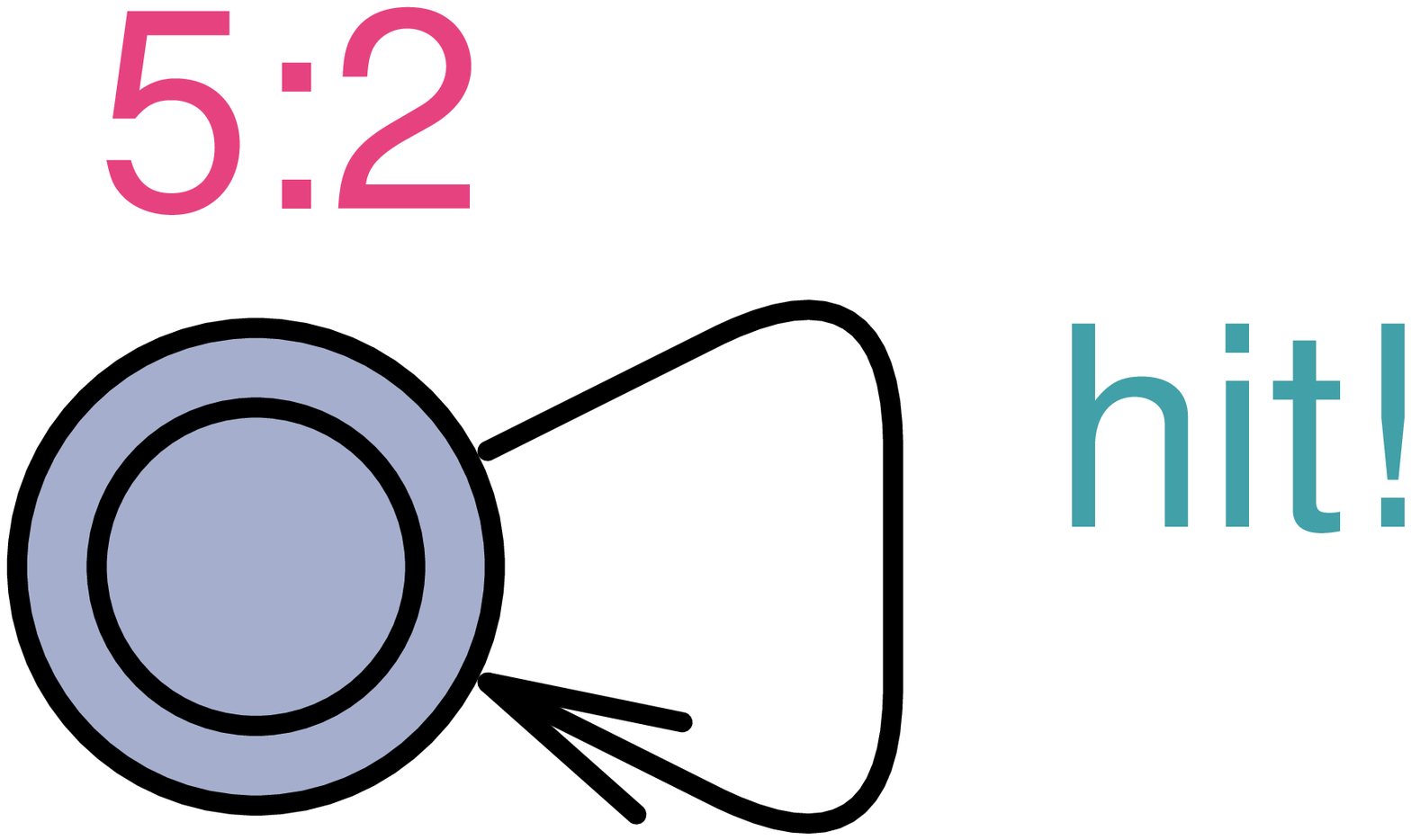}
    \caption{Player  trying to  hit  with exponentially  distributed
      delay.}
    \label{fig:ballpiston}
  \end{subfigure}%
  \begin{subfigure}[b]{0.54\linewidth}
    \includegraphics[width=\linewidth]{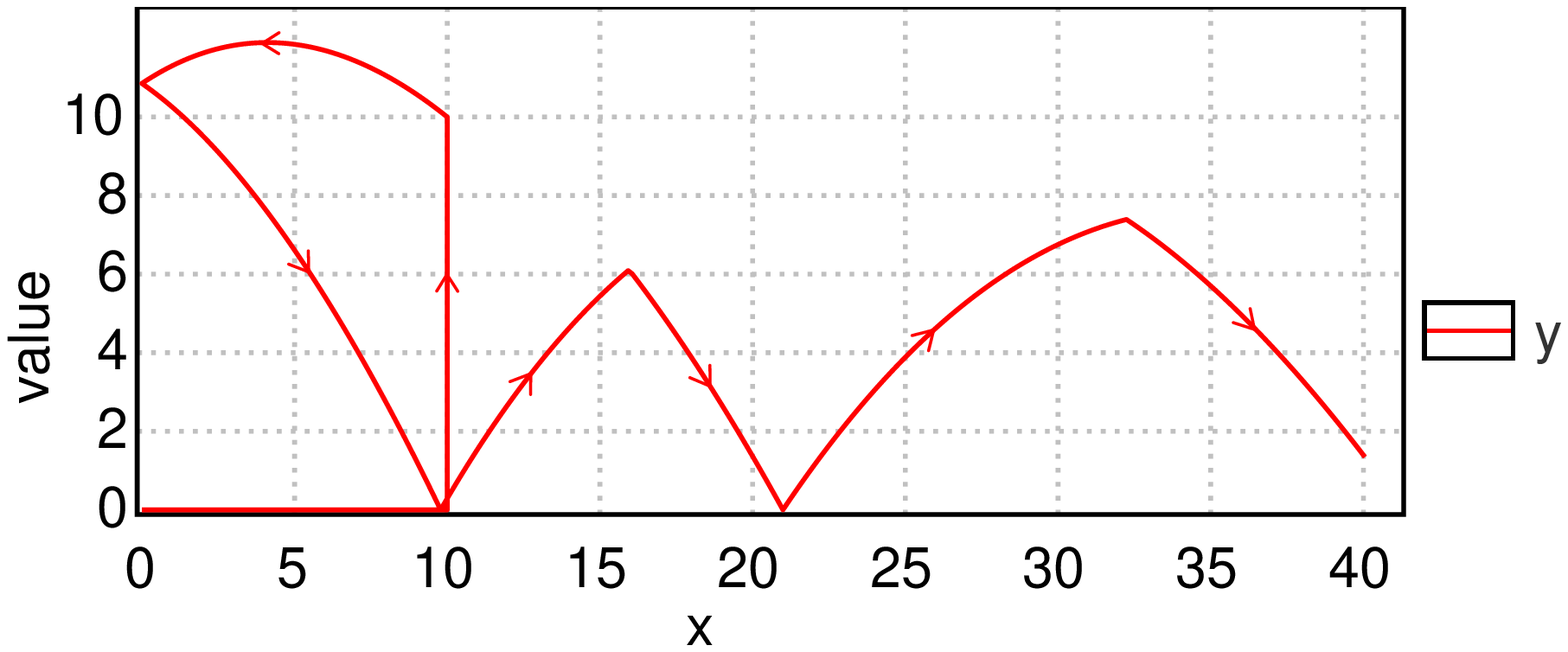}
    \caption{Trajectory as $(x,y)$ plot.}
    \label{fig:ballplot}
  \end{subfigure}
  \caption{Models and a trajectory of a thrown/bouncing ball hit by a player.}
  \label{fig:ball}
\end{figure}
In  our  version\footnote{Our  version  is inspired  by  the  recently
  announced  solution by  a 16  year old  German school  boy (Shouryya
  Ray), to a 350 year open problem by Newton concerned with predicting
  the  trajectory of  a ball  thrown  at a  wall.}
  the  ball is  initially thrown  against a
wall, bounces against it, and then continues its trajectory by falling
and bouncing against the floor.  In addition, a (inexperienced) player
tries  to   hit  the  ball   randomly  according  to   an  exponential
distribution.  The  model is depicted  in Fig.~\ref{fig:ball}(a)--(c).
The  player  is  modelled   as  a  simple  automaton  that  broadcasts
\texttt{hit!}  with  an exponential  distribution of rate  $5/2$.  The
\texttt{x} coordinate (Fig.~\ref{fig:ballx}) is initialised to 10 with
an  uncertain derivative  \texttt{vx}  uniformly between  $[-10,-9.5]$
(the  ball is thrown  against the  wall), after  which the  ball moves
toward  the   wall  (placed  at  0).   Here,   the  automaton  outputs
\texttt{bounce!}   on  an  \emph{urgent}  channel,  which  forces  the
transition to take place deterministically at \texttt{x}$=0$.  After a
bounce  with a  random dampening  factor of  the  velocity \texttt{vx}
uniformly  between  $[0.85,1]$, the  ball  continues  to  move in  the
opposite direction.   The \texttt{y} coordinate (Fig.~\ref{fig:bally})
is initialised  to 10 with  an uncertain derivative  \texttt{vy}.  The
model shows the effect  of gravitation with \texttt{vy'}$=-9.81$.  The
ball bounces  with a random dampening  factor on the floor  (at 0) and
when the ball  is away from the wall (\texttt{x}$\ge  10$) then it can
be hit by  the player provided it is  high enough (\texttt{y}$\ge 6$).
Depending on the current direction of the ball, the ball may bounce or
it  is  pushed.  One  possible  trajectory of  the  ball  is shown  in
Figure~\ref{fig:ballplot}.  The plot is obtained by checking the query
``\texttt{simulate  1 [x<=40]\{y\}}''.   The vertical  line  shows the
ball moving to  its initial position and should  be ignored.  The ball
bounces as  expected against the wall,  the floor, and  the hitting of
the player. \uppaalsmc is able to simulate this hybrid system that has
a  second order  ODE,  a  stochastic controller  (the  player), and  a
stochastic environment (random dampening factor).

In  addition, we may  perform statistical  model-checking in  order to
estimate the probability that the ball is still bouncing above a height
of 4 after 12 time units with the query:
\noindent
\begin{flushleft}
\texttt{Pr[<=20](<> time>=12 and y>=4)}
\end{flushleft}
\noindent which returns the confidence interval
$[0.44,0.55]$ with 95\% confidence after having generated 738 runs.
We can also test for the hypothesis 
\noindent
\begin{flushleft}
\texttt{Pr[<=20](<> time>=12 and y>=4) >= 0.45},
\end{flushleft}
\noindent which gives a more precise lower bound. The hypothesis holds
with a region  of indifference $\pm 0.01$ and  a level of significance
of 5\% after generating 970 runs.

\section{Networks of Hybrid Automata}\label{sec:nha}


\newcommand{\SHA}{SHA\xspace}
\newcommand{\NSHA}{NSHA\xspace}


\uppaalsmc  now supports  the analysis  of stochastic  hybrid automata
(\SHA) that are timed automata whose  clock rates can be changed to be
constants  or  expressions  depending  on  other  clocks,  effectively
defining  ODEs.   This generalizes  the  model  used  in our  previous
work~\cite{DLLMW11,DLLMPVW11} where  only linear priced  automata were
handled.     Our   new    release    \uppaalsmc   4.1.10\footnote{{\tt
    www.uppaal.org}.}  supports fully hybrid  automata with ODEs and a
few built-in  complex functions  (such as {\tt  sin}, {\tt  cos}, {\tt
  log}, {\tt exp} and {\tt sqrt})!

\paragraph{{Hybrid Automata}}
Intuitively, a hybrid automaton ${\cal H}$ is a finite-state automaton
extended with  continuous variables that evolve  according to dynamics
characterizing  each discrete state  (called a  \emph{location}).  Let
$X$  be  a  finite  set  of continuous  variables.   A  \emph{variable
  valuation}  over $X$  is a  mapping $\nu:X\rightarrow  \bbbr$, where
$\bbbr$  is the  set of  reals.   We write  $\bbbr^X$ for  the set  of
valuations over  $X$.  Valuations over $X$ evolve  over time according
to \emph{delay  functions} $F:\rplus\times\bbbr^X\rightarrow \bbbr^X$,
where for a delay $d$ and valuation $\nu$, $F(d,\nu)$ provides the new
valuation after  a delay of $d$.  As  is the case for  delays in timed
automata, delay functions are assumed to be time additive in the sense
that  $F(d_1,F(d_2,\nu))=F(d_1+d_2,\nu)$.  To allow  for communication
between different hybrid automata we assume a set of actions $\Sigma$,
which is partitioned  into disjoint sets of input  and output actions,
i.e. $\Sigma=\Sigma_i\uplus\Sigma_o$.

\begin{definition}
  A Hybrid Automaton (HA) ${\cal  H}$ is a tuple ${\cal H}=(L,\ell_0,X,
  E,  F, I)$,  where:  (i) $L$  is  a finite  set  of locations,  (ii)
  $\ell_0\in L$ is  an initial location, (iii) $X$ is  a finite set of
  continuous  variables,  (iv)  $\Sigma=\Sigma_i\uplus\Sigma_o$  is  a
  finite  set  of actions  partitioned  into  inputs ($\Sigma_i$)  and
  outputs ($\Sigma_o$), (v)  $E$ is a finite set of  edges of the form
  $(\ell, g, a, \phi, \ell')$, where $\ell$ and $\ell'$ are locations,
  $g$ is a  predicate on $\bbbr^X$, action label $a\in\Sigma$ and 
  $\phi$ is  a binary relation on $\bbbr^X$,  (vi)  for  each  location
  $\ell\in L$  $F(\ell)$  is  a \emph{delay function},  and (vii) $I$ 
  assigns  an invariant predicate $I(\ell)$ to any location $\ell$.
\end{definition}


The  semantics of  a  HA ${\cal  H}$  is a  timed labeled  transition
system,  whose states  are pairs  $(\ell,\nu)\in  L\times\bbbr^X$ with
$\nu\models   I(\ell)$,  and  whose   transitions  are   either  delay
transitions  $(\ell,\nu)\arrow{d}(\ell,\nu')$  with  $d\in\rplus$  and
$\nu'=F(d,\nu)$,           or           discrete           transitions
$(\ell,\nu)\arrow{a}(\ell',\nu')$     if    there    is     an    edge
$(\ell,g,a,\phi,\ell')$ such that $\nu\models g$ and $\phi(\nu,\nu')$.
We  write  $(\ell,\nu)\leadsto(\ell',\nu')$   if  there  is  a  finite
sequence  of  delay  and  discrete transitions  from  $(\ell,\nu)$  to
$(\ell',\nu')$.

In the above  definition, we have deliberately left  open the concrete
syntax for  the delay function  $F$ as well  as the invariant  $I$. In
\uppaalsmc the delay update for a simple clock $x$ -- used in (priced)
timed automata --  is given by an implicit rate  $x'=1$ or an explicit
rate  $x'=e$ appearing in  the invariant  of $\ell$,  where $e$  is an
expression only depending  on the discrete part of  the current state.
More generally, the effect of  the delay function $F$ may be specified
by a set of ODEs needing to be solved. It is important to note that in
specifying  the delay  function $F$  and the  invariant $I$,  the full
syntax of  \uppaal expressions -- including  user-defined functions --
is at the disposal.

\begin{example} 
  Reconsider   the  extended  bouncing   ball  example   from  Section
  \ref{sec:ball}. Here the automaton for the \texttt{x} coordinate may
  be  initialized  to  the state  $(\mathtt{x}=10,\mathtt{vx}=-9.8)$,
  after which the following transition sequence may occur:
\[
(\mathtt{x}=10,\mathtt{vx}=-9.8)
\arrow{10\div 9.8} (\mathtt{x}=0,\mathtt{vx}=-9.8)
\arrow{\mathtt{bounce!}} (\mathtt{x}=0,\mathtt{vx}=9.31)
\]
where  in  the \texttt{bounce!}-transition  the  dampening factor  has
non-deterministically been chosen  from the interval $[0.85,1.00]$ as
$0.95$.   The  automaton   for  the   \texttt{y}  coordinate   may  be
initialized to  the state $(\mathtt{y}=10,\mathtt{vy}=  5.9)$ after
which  the   delay  function  will  effectively  be   given  by  delay
transitions:
\[
(\mathtt{y}=10,\mathtt{vy}=5.9)\arrow{d}  (\mathtt{y}=-9.81/2 d^2 +
5.9 d + 10,\mathtt{vy}=-9.81 d + 5.9)
\]
The ball \texttt{bounce!}'es on the floor, i.e. $\mathtt{y}=0$ or when
(approximately)    $d=2.15$,   and    will    afterwards   -    having
non-deterministically  chosen the  dampening factor  $0.9$  - continue
from the state $(\mathtt{y}=0,\mathtt{vy}=13.67)$.
\end{example}

\paragraph{{Networks of Hybrid Automata}}
Following the compositional specification  theory for timed systems in
\cite{hscc2010},    we    shall    assume    that   NHAs    are 
\emph{input-enabled} in  the sense,  that for all  states $(\ell,\nu)$
and input  actions $\iota\in\Sigma_i$,  for all HAs  $j$, there  is an
edge  $(\ell^j,g,\iota,\phi,{\ell^j}')$ such  that $\nu\models  g$ and
$\phi(\nu,\nu')$  for some  valuation $\nu'$.   Also, we  shall assume
that time always diverges,  and that different automata synchronize on
matching    inputs    and   outputs    as    a   standard    broadcast
synchronization~\cite{gomez09}.
%

Whenever  ${\cal A}^j=(L^j,X^j,\Sigma^j,E^j,F^j,I^j)$  ($j=1\ldots n$)
are NHA,  they are \emph{composable} into a  \emph{closed network} iff
their variable-sets are  disjoint ($X^j\cap X^k=\emptyset$ when $j\neq
k$), they have the same action set ($\Sigma=\Sigma^j=\Sigma^k$ for all
$j,k$), and  their output action-sets provide a  partition of $\Sigma$
($\Sigma^j_o\cap\Sigma^k_o=\emptyset$     for    $j\neq     k$,    and
$\Sigma=\cup_j \Sigma^j_o$).  For $a\in\Sigma$ we denote by $c(a)$ the
unique $j$ with $a\in\Sigma^j$. If $\nu\in\bbbr^X$ with $X=\cup_j X^j$, we denote by $\nu\downarrow_{X^j}$ the projection of $\nu$ to $X^j$.

\begin{definition}
  Let  ${\cal A}^j=(L^j,X^j,\Sigma,E^j,F^j,I^j)$ (with  $j=1\ldots n$)
  be    composable    NHAs.     Their    \emph{composition}    $({\cal
    A}_1\,|\ldots|\,{\cal   A}_n)$   is   the  HA   $\boldsymbol{{\cal
      A}}=(L,X,\Sigma,E,F,I)$   where    (i)   $L=\times_jL^j$,   (ii)
  $X=\cup_j                         X^j$,                        (iii)
  $F(\boldsymbol{\ell})(d,\nu)(x)=F^j(\ell^j)(d,\nu\downarrow_{X^j})(x)$
  when $x\in  X^j$, (iv) $I(\boldsymbol{\ell})=\cap_j  I(\ell^j)$, and
  (v)              $(\boldsymbol{\ell},\cap_j             g_j,a,\cup_j
  \phi_j,\boldsymbol{\ell'})\in               E$              whenever
  $(\ell_j,g_j,a,\phi_j,\ell'_j)\in E^j$ for $j=1\ldots n$.
\end{definition}

\section{Stochastic Semantics for Networks of Hybrid Automata}
\label{sec:sha}
Reconsidering  again our extended  version of  the bouncing  ball from
Section  \ref{sec:ball}, it  is clear  that there  is a  constant race
between  the ball  \texttt{bounce!}ing  on the  floor  and the  player
\texttt{hit!}ing  the   ball.   Whereas   the  time  of   bouncing  is
deterministic -- given by the ODE obtained from the (stochastic) effect
of  the previous \texttt{bounce!}   or \texttt{hit!}   -- the  time of
hitting is  stochastic according  to an exponential  distribution with
rate $5/2$.   However based on this, a  measure on sets of  runs of the
systems is induced, according to which quantitative properties such as
\emph{``the probability that the ball with have a height greater than 4
  after 12 time-units''} become well-defined.



Our  early  works  \cite{DLLMPVW11}  --  though  aimed  at  stochastic
semantics of  \emph{priced timed automata} --  is sufficiently general
that it also provides the basis for a natural stochastic semantics for
networks of HAs,  where components associate probability distributions
to  both the time-delays  spent in  a given  state as  well as  to the
transition between states.

Let  ${\cal A}^j=(L^j,X^j,\Sigma,E^j,F^j,I^j)$  ($j=1\ldots  n$) be  a
collection   of    composable   HAs.    Under    the   assumption   of
input-enabledness,  disjointedness  of clock  sets  and output  actions,
states   of   the    composite   NHA   $\boldsymbol{{\cal   A}}=({\cal
  A}_1\,|\ldots|\,{\cal  A}_n)$  may  be  seen  as  tuples  ${\bf  s}=
(s_1,\ldots,s_n)$ where $s_j$ is a state of ${\cal A}^j$, i.e.  of the
form  $(\ell,\nu)$ where $\ell\in  L^j$ and  $\nu\in\bbbr^{X^j}$.  Our
probabilistic  semantics is  based  on the  principle of  independence
between components.   Repeatedly each component decides on  its own --
based  on  a  given  delay  density function  and  output  probability
function --  how much  to delay before  outputting and what  output to
broadcast  at  that  moment.    Obviously,  in  such  a  race  between
components the  outcome will be  determined by the component  that has
chosen to output after the  minimum delay: the output is broadcast and
all other components may consequently change state.

\paragraph{{Stochastic Semantics of HA Components}}
The stochastic  semantics of HAs refine  the non-deterministic choices
that  may exist  with respect  to delay,  output and  next  state.  We
consider  the  component  ${\cal  A}^j$  and let  $\St^j$  denote  the
corresponding set of states.   For each state $s=(\ell,\nu)$ of ${\cal
  A}^j$, we  shall assume  that there exist  probability distributions
for  delays, output as  well as  next-state:  
\begin{itemize}
\item  the  \emph{delay density  function},  $\mu_s$  over delays  in
  $\rplus$,  provides stochastic  information for  when  the component
  will perform an output, thus $\int \mu_s(t) dt=1$;
\item  the  \emph{output   probability  function}  $\gamma_s$  assigns
  probabilities   for  resolving   what  output   $o\in\Sigma^j_o$  to
  generate, i.e. $\sum_o \gamma_s(o) =1$;
\item  the  \emph{next-state   density  function}  $\eta_s^a$  provide
  stochastic information on the next state $s'=(\ell',\nu')\in\bbbr^X$
  given an action $a$, i.e. $\int_{s'}\eta_s^a(s') =1$.
\end{itemize}
For outputs happening deterministically at an exact time point $d$ (or
deterministic next states $s'$), $\mu_s$ ($\eta_s^a$) becomes a Dirac delta
function $\delta_d$ ($\delta_{s'}$)\footnote{which should  formally be treated as the
  limit  of a  sequence of  delay density  functions  with decreasing,
  non-zero support around $d$.}.
%
%

In \uppaalsmc uniform distributions are applied for states where delay
is  bounded, and  exponential  distributions (with  location-specified
rates)  are  applied for  the  cases,  where  a component  can  remain
indefinitely  in a  location.   Also, \uppaalsmc  provides syntax  for
assigning  discrete  probabilities to  different  outputs  as well  as
specifying stochastic distributions on next-states (using the function
{\tt random[b]} denoting a uniform distribution on $[0,b]$).

\paragraph{{Stochastic Semantics of Networks of HA}}
%
%
For  the  stochastic  semantics  of  closed networks  of  HA  consider
$\boldsymbol{{\cal  A}}=({\cal  A}_1\,|\ldots|\,{\cal  A}_n)$  with  a
state  space $\St=\St_1\times\cdots\times\St_n$.   For  ${\bf s}=(s_1,
\ldots,s_n)\in\St$  and $a_1a_2\ldots  a_k\in\Sigma^{*}$ we  denote by
$\pi({\bf s},a_1a_2\ldots a_k)$ the set of all maximal runs from ${\bf
  s}$   with   a   prefix   $t_1a_1t_2a_2\ldots   t_ka_k$   for   some
$t_1,\ldots,t_n\in\rplus$, that is runs  where the $i$'th action $a_i$
has been outputted by the component $A_{c(a_i)}$.  Providing the basic
elements of  a Sigma-algebra, we now inductively  define the following
measure for such sets of runs:
\begin{small}
\begin{eqnarray*}
\lefteqn{\P_{\A}\big(\pi({\bf s},a_1\ldots a_n)\big)  =} \\
 & &  \int_{t\geq 0}
    \hspace{-1pt} \mu_{s_c}(t) \cdot 
    \big(\prod_{j\not=c}\int_{\tau>t} \hspace{-1pt} \mu_{s_j}(\tau)
    d\tau \big) 
    \cdot 
   \gamma_{{s_c}^t}(a_1) \cdot 
   \int_{\bf s'} \big(\prod_j\eta_{s_j^t}^{a_1}(s'_j) \cdot
    \P_{\A}\big(\pi({\bf s'},a_2\ldots a_n)\big) d{\bf s'}\big)\, dt
\end{eqnarray*}
\end{small}

This definition requires a few  words of explanation: at the outermost
level we integrate over all  possible initial delays $t$.  For a given
delay $t$, the outputting component $c=c(a_1)$ will choose to make the
broadcast  at time $t$  with the  stated density.   Independently, the
other components will choose to a  delay amount, which -- in order for
$c$ to be the winner -- must  be larger than $t$; hence the product of
the probabilities that  they each make such a  choice.  Having decided
for  making the  broadcast at  time $t$,  the probability  of actually
outputting $a_1$  is included.   Finally, integrating over  all global
states ${\bf s'}$  that may result from all  components having delayed
$t$ time-units  and changed state  stochastically with respect  to the
broadcasted  action $a_1$,  the probability  of runs  according  to the
remaining actions $a_2\ldots a_n$ is taken into account.





\paragraph{{Generation of Random Runs}}
The central component of the SMC engine of \uppaalsmc is the efficient
generation  of  random  runs  according to  the  stochastic  semantics
proposed in the preceding Section.
%
%
\uppaalsmc has  to integrate the ODEs  given in the  model while still
respecting  its stochastic  semantics.  The  tool does  not  solve the
equations exactly and currently supports the Euler integration method.
A  fixed time step  $\delta t$  (defined by  the user)  is used  by an
internal integrator  component added  to the system.   This integrator
races with the other processes during  its time step and all rates are
considered constant and defined by the equations in the model.  At the
end of every  step or if another process wins the  race, all the rates
are re-evaluated and  the resulting values of the  clocks are computed
according to  $x_{new} = x_{old} +  \delta t * x_{old}'$.   We plan to
implement  more  robust  methods,  such as  Runge-Kutta's,  that  give
different evaluations  of $x_{new}$. In this case,  $x_{old}'$ is then
derived  from the method  and not  directly from  the equation  in the
model and  kept constant during $\delta t$.   The stochastic semantics
and the races between the components are still respected.

The integrator component always picks  a delay equal to the given time
step $\delta t$.  The other  components pick delays according to their
distributions. The  winning component delays  to the point in  time it
wanted and  tries to take  a transition at  that point. If  it appears
that no guard is satisfied,  no transition is taken and all components
race again. In practice, every component will use the current rates of
the clocks to evaluate a lower  bound on the delay from which anything
can happen to skip delaying for nothing.

\paragraph{{Statistical Model Checking}}

We use  SMC\,\cite{LS91,SVA04,YS02,LDB10} to estimate and  test on the
probability that  a random  run of  a network of  SHAs will  satisfy a
given property.  Given a model  ${\cal H}$ and a trace property $\phi$
(e.g. expressed in LTL \cite{LTL}  or MTL \cite{MTL}), SMC refers to a
series of simulation-based  techniques that can be used  to answer two
questions: (1)  \emph{Qualitative:} is  the probability that  a random
run of  ${\cal H}$ will satisfy  $\phi$ greater or equal  to a certain
threshold $\theta$ (or greater or  equal to the probability to satisfy
another property  $\phi'$)?  and (2) \emph{Quantitative:}  what is the
probability that a  random run of ${\cal H}$  will satisfy $\phi$?  In
both cases, the answer will be correct up to a user-specified level of
confidence,  providing  a upper  bound  on  the  probability that  the
conclusion made by the algorithm  will be wrong.  For the quantitative
approach,  which we  will use  intensively in  this paper,  the method
computes a  confidence interval that  is an interval  of probabilities
that contains the true probability  to satisfy the property.  Here the
confidence level provides the probability that the computed confidence
interval indeed contains the unknown probability.

Our \uppaalsmc tool-set  implements a wide range of  SMC algorithms for
networks of SHAs not only  for reachability and safety properties, but
also  for  general  weighted  MTL properties  \cite{MITLSMC,WMTL}.  In
addition, the tool offers several  features to visualize and reason on
the results.

\section{Energy Aware Buildings}\label{build-sec}

\uppaalsmc  has  recently   \cite{chinesepaper}  been  applied  to  an
evaluation framework for energy  aware buildings, where the control of
heating is optimized with  respect to environmental and user profiles.
To indicate central  parts of this framework and  the clear benefit of
modeling  with  ODEs, we  illustrate  the  framework  on a  simplified
instance  and recall results  from our  case study~\cite{chinesepaper}
based on a benchmark for hybrid systems verification \cite{Fehnker04}.

\begin{figure}[tb]
\centering
\begin{tabular}{ccc}
\includegraphics[width=0.24\linewidth]{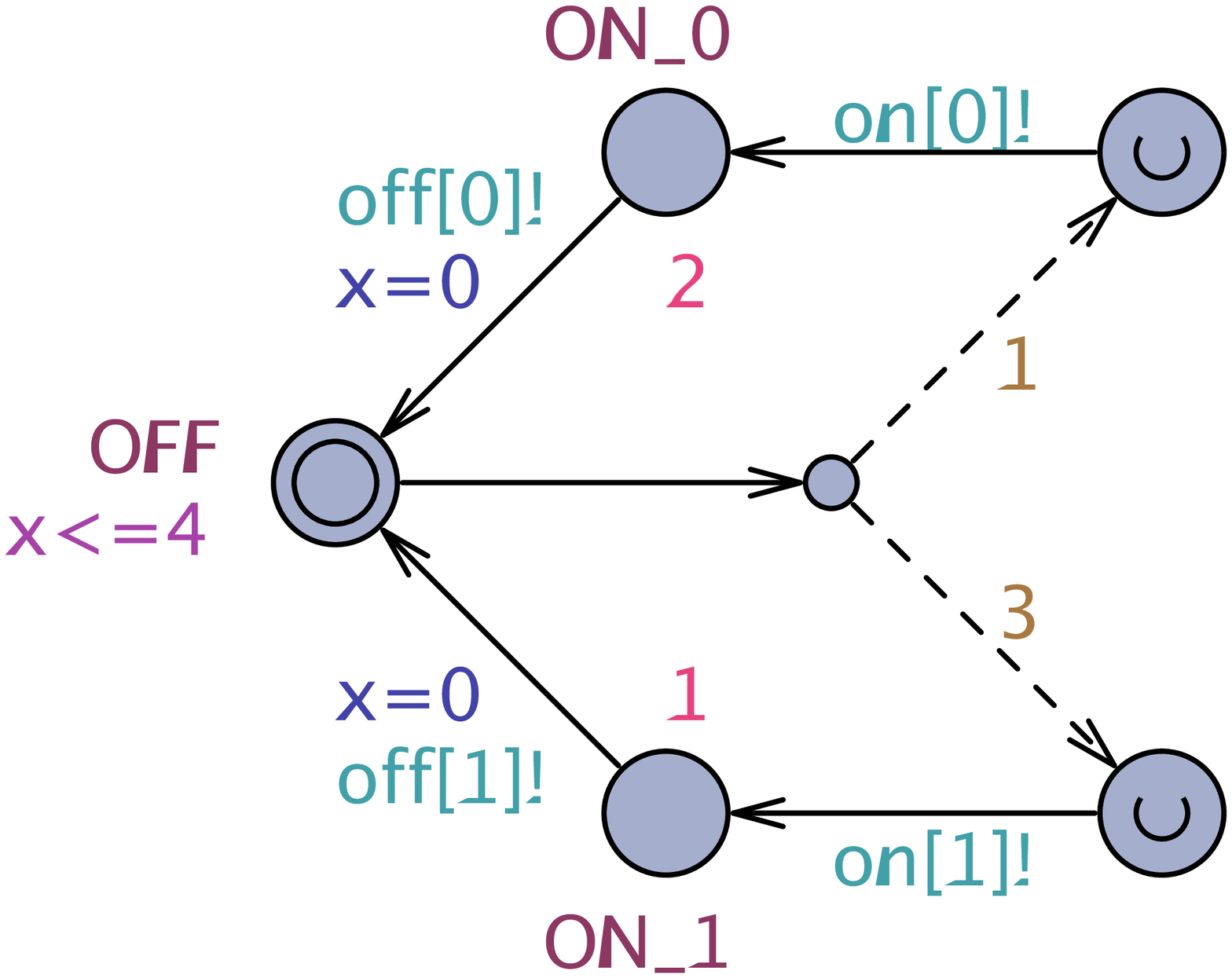} &
\includegraphics[width=0.35\linewidth]{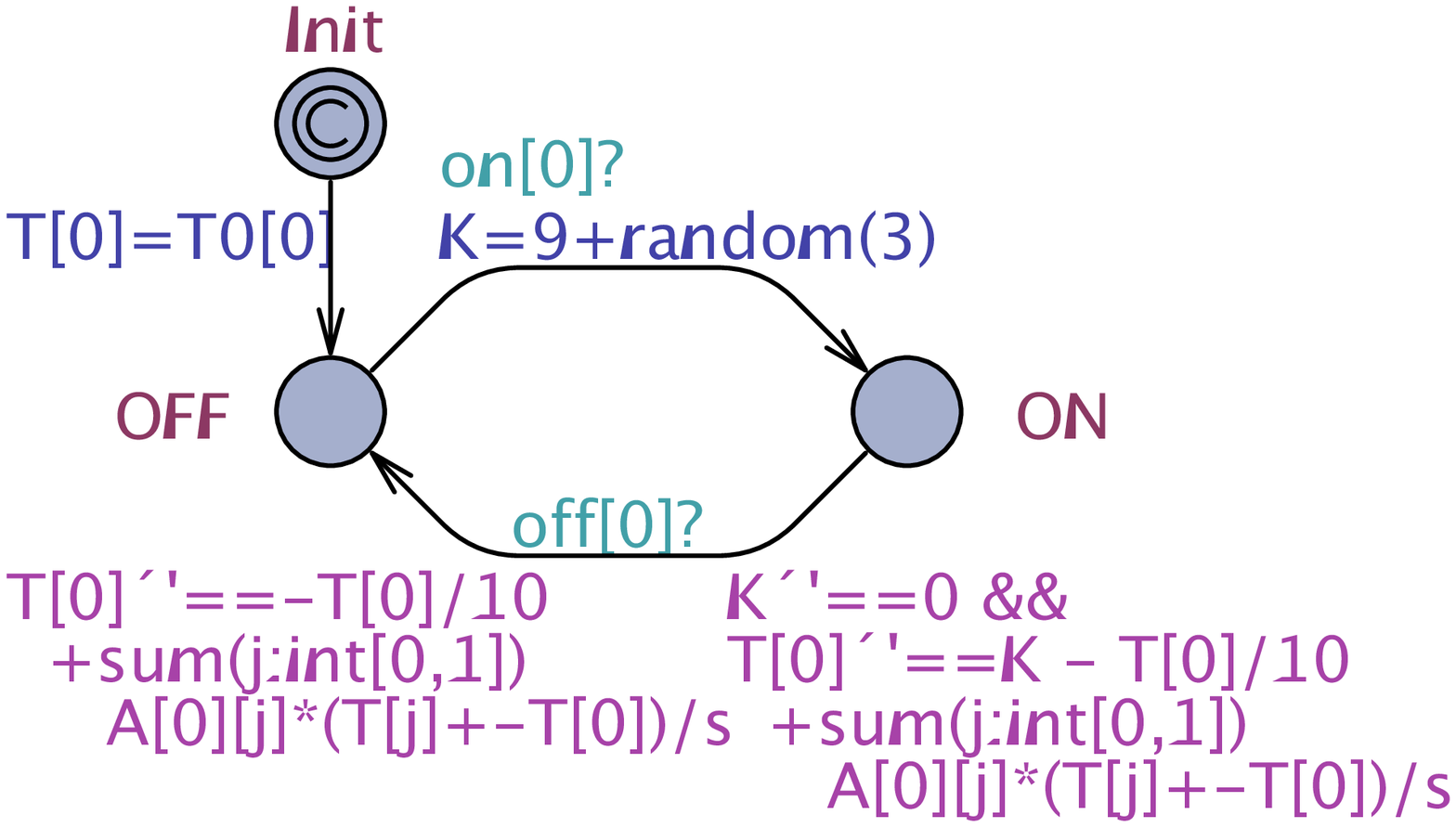} &
\hspace{-10pt}
\includegraphics[width=0.35\linewidth]{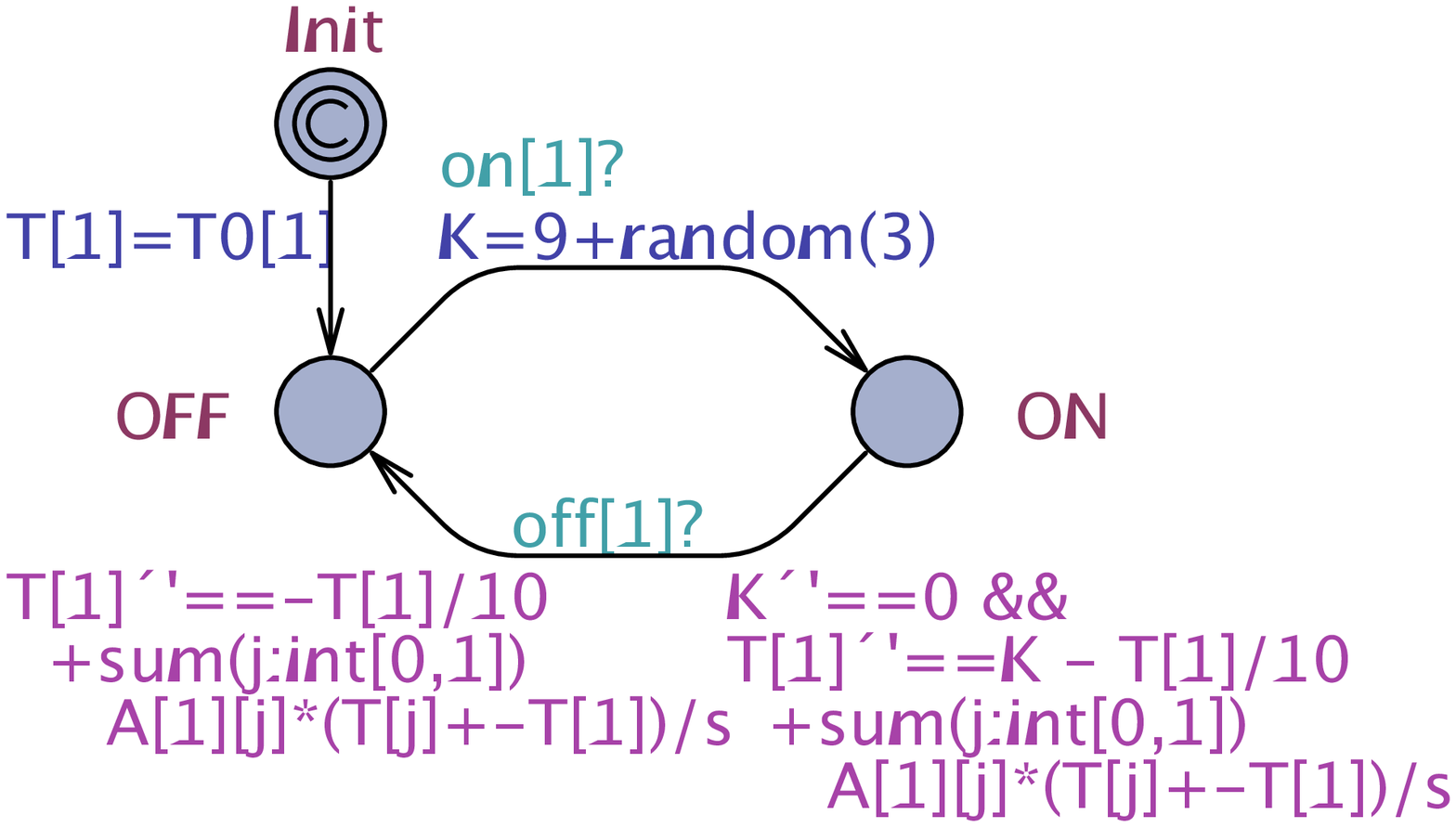} \\
(a) stochastic heater. & (b) room 0. & (c) room 1.
\end{tabular}
\vspace*{-3mm}
\caption{A simple two room example with an autonomous heater.}
\label{fig:2rooms1}
\vspace*{-2mm}
\end{figure}

\paragraph{A Simple 2-Room Example}
To illustrate the various aspects of the (extended) modeling formalism
supported by \uppaalsmc, we consider the case of two independent rooms
that can be  heated by a single heater shared by  the two rooms, i.e.,
at most one room can be heated at a time.  Fig.\ref{fig:2rooms1}(a)
shows the automaton for the heater.  It turns itself on with a uniform
distribution   over  time   in-between  $[0,4]$   time   units.   With
probability $1/4$ room 0 is  chosen and with probability $3/4$ room 1.
The heater stays on for some time given by an exponential distribution
(rate 2 for room  0, rate 1 for the room 1).   In summary, one may say
that the  controller is more eager  to initiate the heating  of room 1
than room 0, as well as less  eager to stop heating room 1.  The rooms
are similar and are modeled by the same template instantiated twice as
shown  in Fig.~\ref{fig:2rooms1}(b-c).  The  room is  initialized to
its initial  temperature and then  depending on whether the  heater is
turned  on  or not,  the  evolution of  the  temperature  is given  by
$T_i'=-T_i/10+\sum_{j=0,1}A_{i,j}(T_j-T_i)$                          or
$T_i'=K-T_i/10+\sum_{j=0,1}A_{i,j}(T_j-T_i)$ where  $i,j=0,1$ are room
identifiers.  The sum expression corresponds to an energy flow between
rooms and  matrix $A$ encodes the energy  transfer coefficient between
adjacent  rooms.   Furthermore, when  the  heater  is  turned on,  its
heating is not  exact and is picked with a  uniform distribution of $K
\in [9,12]$, realized by the update \texttt{K=9+random(3)}.

This  example illustrates  the support  for \NSHA  in  \uppaalsmc with
extended arithmetics on clocks and generalized clock rates.

\paragraph{Extended Input Language}
\uppaalsmc  takes as  input \NSHA  as described  above.  Additionally,
there is  support for  other features of  the \uppaal  model checker's
input  language  such  as   integer  variables,  data  structures  and
user-defined  functions,  which  greatly ease  modeling.   \uppaalsmc
allows the user to specify arbitrary rates for the clocks, which includes
a mix of integer and clock expressions on  any location.
In  addition, the  automata support  branching edges
where  weights  can  be  added  to give  a  distribution  on  discrete
transitions. It  is important  to note that  rates and weights  may be
general  expressions that  depend on  the states  and not  just simple
constants.

\paragraph{Checking Queries}
The fundamental principle in \uppaalsmc is to generate runs and evaluate
some expression on the states along the obtained run. Runs are always
\emph{bounded}, either by time, by a number of steps, or more generally
by cost (when using a clock explicitly). The engine has a built-in
heuristic detection of Zeno behaviours to abort the generation of such runs.
Examples of the syntax for the different types of bounds are \texttt{[<=100]}
for 100 time units since the beginning of the run, \texttt{[\#<=50]} for
50 discrete transitions taken from the initial state, and \texttt{[x<=200]}
until the clock \texttt{x} reaches 200\footnote{It is up to the modeler to
ensure that the clock eventually reaches the bound.}.

\begin{figure}[t]
  \centering
  \includegraphics[width=\linewidth]{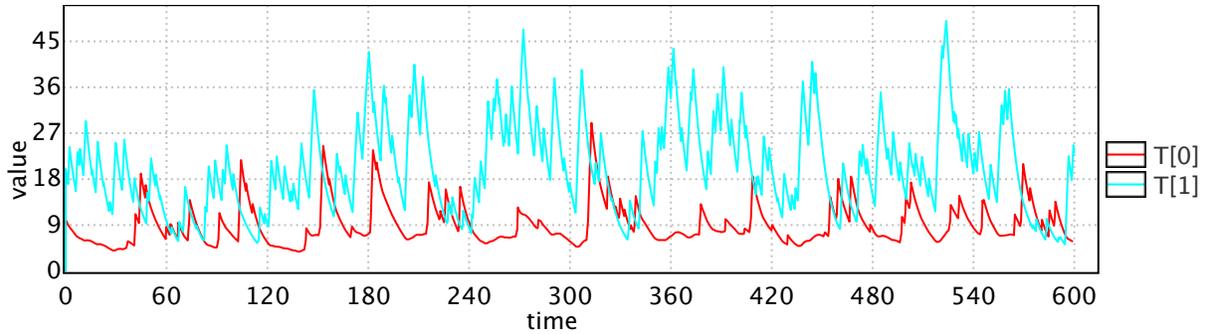}
  \vspace*{-7mm}
  \caption{Evolution of the temperatures of the two rooms.}
  \label{fig:2rooms2}
  \vspace*{-2mm}
\end{figure}

\uppaalsmc supports simulations with monitoring custom expressions,
probability evaluation, hypothesis testing, and probability comparison.
We can simulate and plot the temperatures with the query
\noindent
\begin{flushleft}
\texttt{simulate 1 [<=600]\{T[0],T[1]\}}
\end{flushleft}
The query  request the  checker to provide  one simulate run  over 600
time  units   and  plot  the  temperatures   of  \texttt{Room(0)}  and
\texttt{Room(1)}.  The heater in this example is purely stochastic and
is  not  intended  to  enforce  any  particular  property.   Yet,  the
simulation  obtained from this  query in  Fig.~\ref{fig:2rooms2} shows
that the heater  is able to maintain the  temperatures within (mostly)
distinct intervals.

We can evaluate on a shorter time scale the probability for the
temperature of \texttt{Room(0)} to stay below 30 and the temperature of
\texttt{Room(1)} to stay above 5 with the queries
\noindent
\begin{flushleft}
\texttt{Pr[<=100]([] Room(0).Init || T[0] <= 20)\\
Pr[<=100]([] Room(1).Init || T[1] >= 7)}
\end{flushleft}
The results are respectively in $[0.45,0.55]$ and $[0.65,0.75]$.
The precision and confidence of these so-called confidence intervals
 are user-defined and influence the number
of runs needed to compute the probability. In this example, for
having the precision to be $\pm 0.05$ with a confidence of
95\%, we need 738 runs.
In fact if we are only interested in knowing if the second probability
is above a threshold it may be more efficient to test the hypothesis
\noindent
\begin{flushleft}
\texttt{Pr[<=100]([] Room(1).Init || T[1] >= 7) >= 0.69}
\end{flushleft}
which is accepted in our case with 902 runs for a level of significance
of 95\%. To obtain an answer at comparable level of precision with probability
evaluation, we would need to use a precision of $\pm 0.005$, which
would require 73778 runs instead.

The tool can also compare probabilities without needing to compute them
individually. We can test the hypothesis that the heater is better at keeping the
temperature of \texttt{Room(1)} above 8 than keeping the temperature of
\texttt{Room(0)} below 20:
\noindent
\begin{flushleft}
\texttt{Pr[<=100]([] Room(1).Init || T[1] >= 7) >=\\
Pr[<=100]([] Room(0).Init || T[0] <= 20)}
\end{flushleft}
which is accepted in this case with 95\% level of significance with
just 258 runs.

\begin{figure}[!htb]
  \vspace{-5mm}
  \begin{subfigure}{0.32\linewidth}
    \includegraphics[height=0.15\textheight]{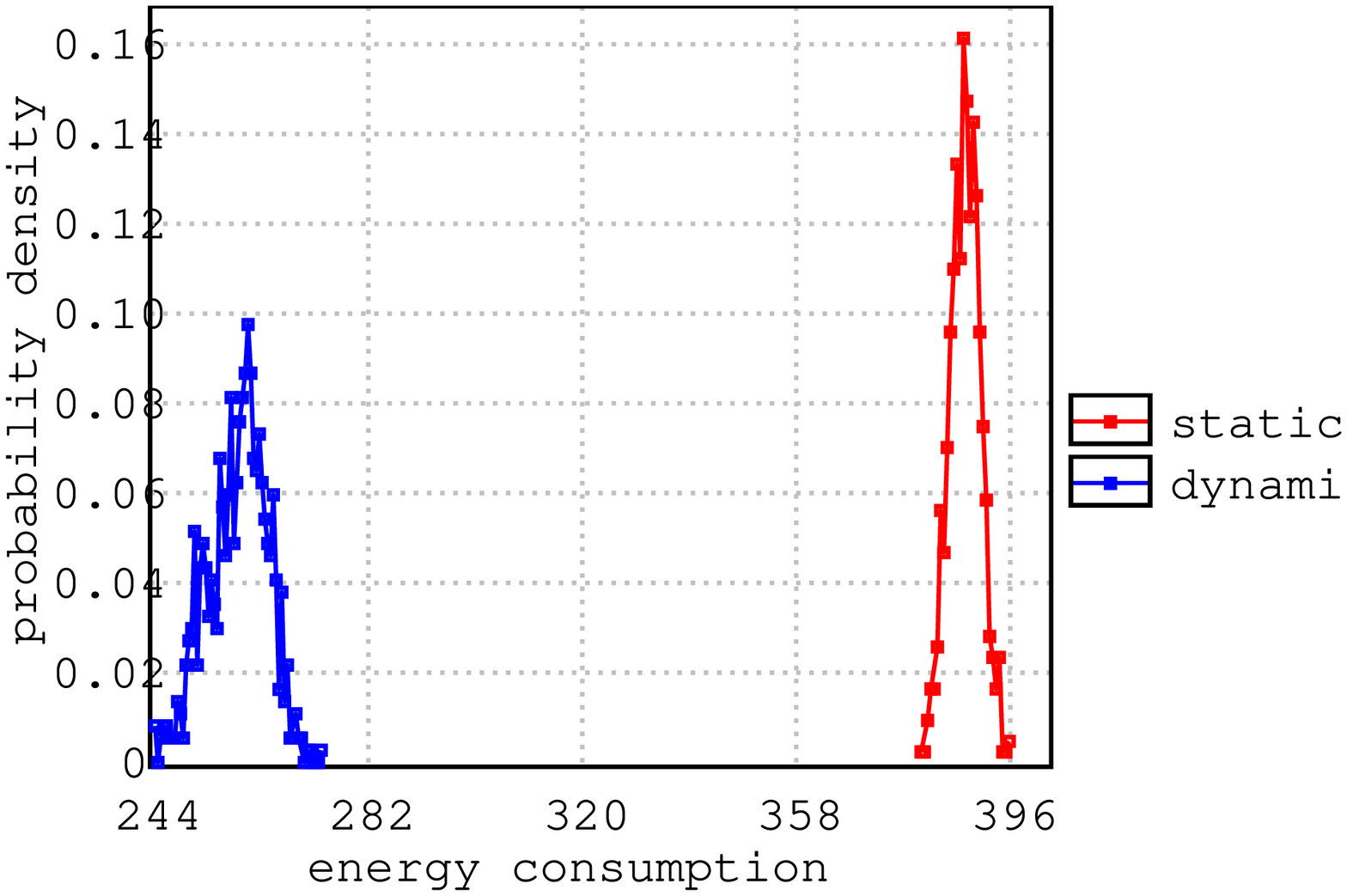}
    \caption{Strategy 1.}
    \label{fig:energyu1}
  \end{subfigure}
  \begin{subfigure}{0.32\linewidth}
    \includegraphics[height=0.15\textheight]{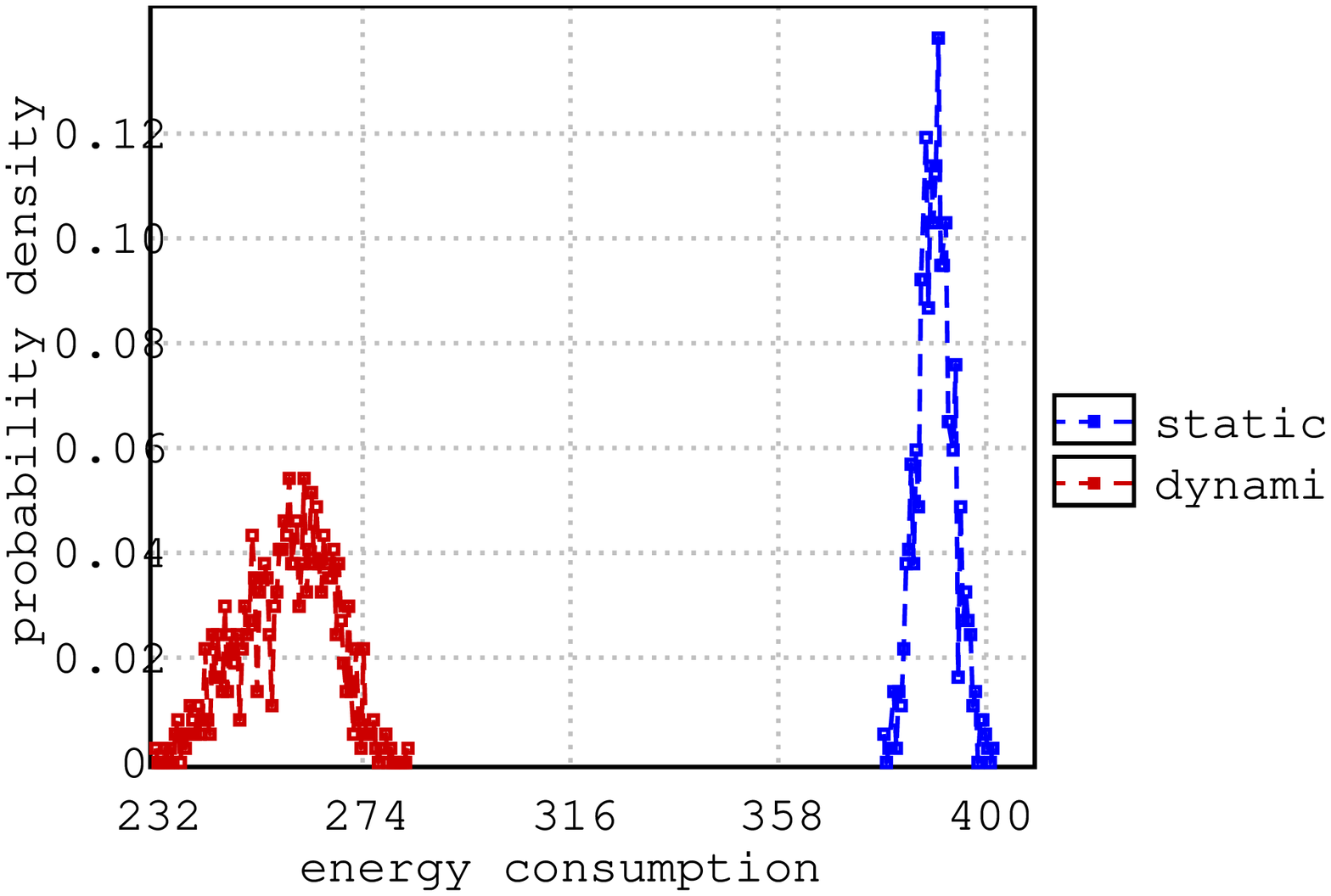}
    \caption{Strategy 2.}
    \label{fig:energyu2}
  \end{subfigure}
  \hspace{\stretch{1}}
  \begin{subfigure}{0.32\linewidth}
    \includegraphics[height=0.15\textheight]{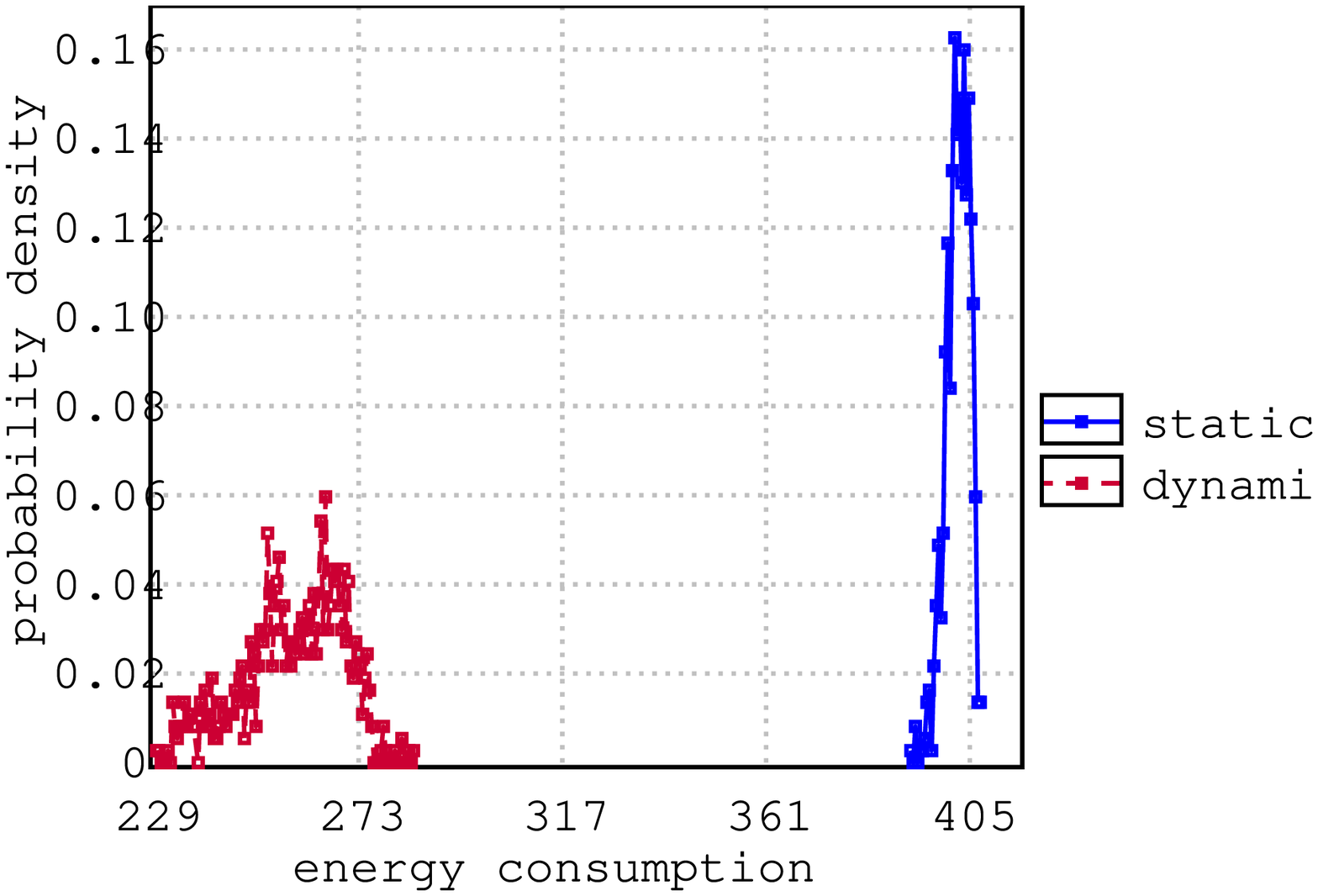}
    \caption{Strategy 3.}
    \label{fig:energyu3}
  \end{subfigure}
  \caption{Energy consumption estimates for static and dynamic user profiles.}
  \label{fig:userdiff}
\end{figure}

\paragraph{Results}
In~\cite{chinesepaper}  we have estimated  the comfort  time (duration
while being  in comfortable temperature range)  and energy consumption
for various  weather conditions, user profiles  and central controller
strategies.   Fig.~\ref{fig:userdiff}  shows  six  energy  consumption
estimates in different  configurations (a building with 5  rooms and 3
heaters).  The  energy comparison shows that the  dynamic user profile
can save  more than  33\% of energy  regardless of the  chosen central
controller strategy.

\section{Biological Oscillator}
\label{bio-sec}
\begin{figure}[ht]
  \begin{subfigure}[b]{0.45\textwidth}
{\scriptsize
\begin{eqnarray*}
dD_A/dt & = &\theta_AD'_A-\gamma_AD_AA\\
dD_R/dt & = &\theta_RD_R'-\gamma_RD_RA\\
dD_A'/dt & = &\gamma_AD_AA-\theta_AD_A'\\
dD_R'/dt & = &\gamma_RD_RA-\theta_RD_R'\\
dM_A/dt & = &\alpha_A'D_A'+\alpha_AD_A-\delta_{M_A}M_A\\
dM_R/dt & = &\alpha_R'D_R'+\alpha_RD_R-\delta_{M_R}M_R\\
dA/dt & = &\beta_AM_A+\theta_AD_A'+\theta_RD_R'\\&-&A(\gamma_AD_A+\gamma_RD_R+\gamma_CR+\delta_A)\nonumber\\
dR/dt & = &\beta_RM_R-\gamma_CAR+\delta_AC-\delta_RR\\
dC/dt & = &\gamma_CAR-\delta_AC
\end{eqnarray*}}
\vspace*{-6mm}
\caption{Ordinary differential equations.}
\label{fig:ODE}
  \end{subfigure}
  \begin{subfigure}[b]{0.5\textwidth}
    \includegraphics[trim=0mm 0mm 20mm 0mm,clip,width=\linewidth]{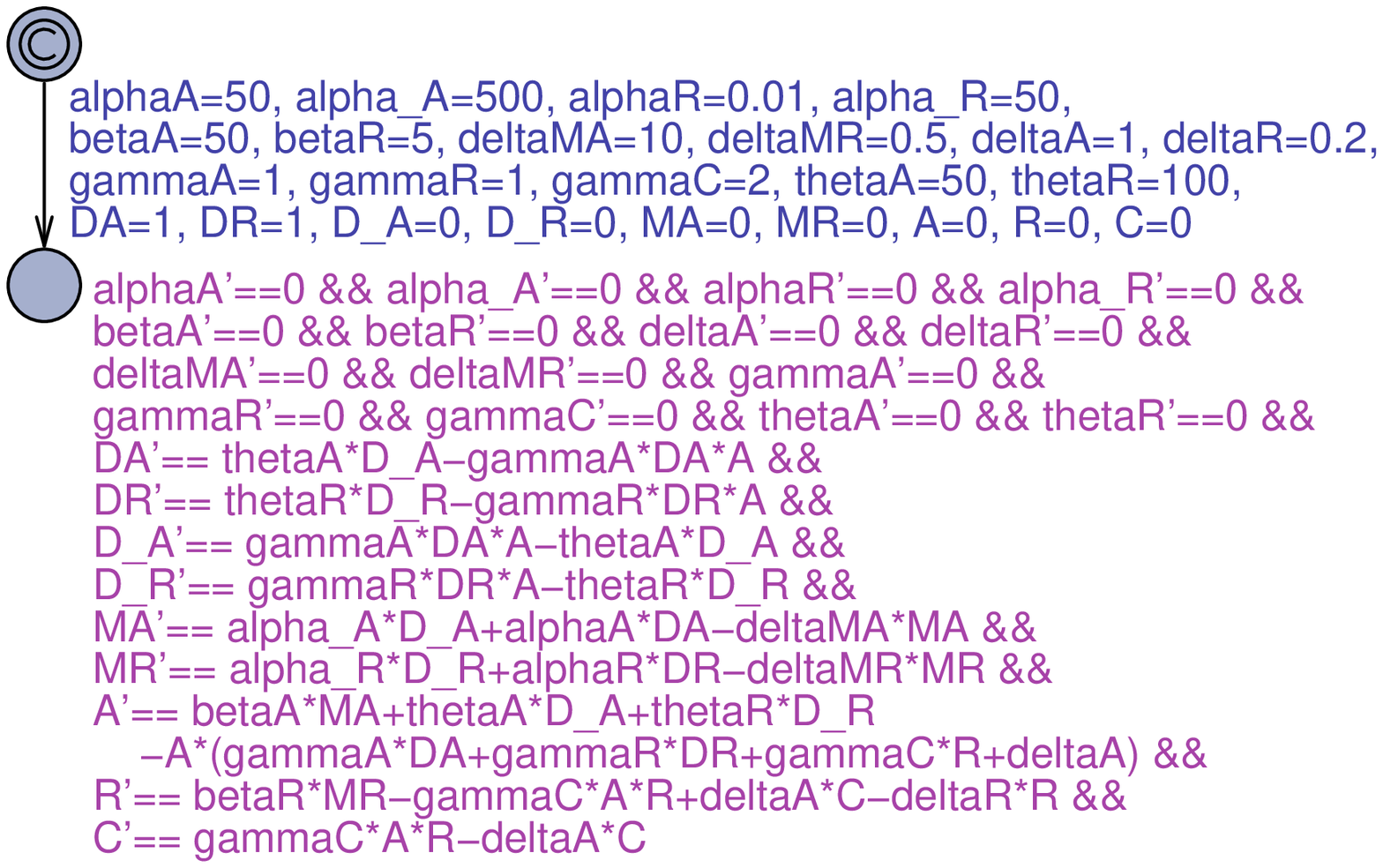}
    \caption{\uppaal automaton representation.}
    \label{fig:uppaalode}
  \end{subfigure}
  \caption{Dynamics of genetic oscillator.}
\end{figure}

\begin{figure}[!htb]
  \begin{subfigure}{\linewidth}
    \includegraphics[width=\linewidth]{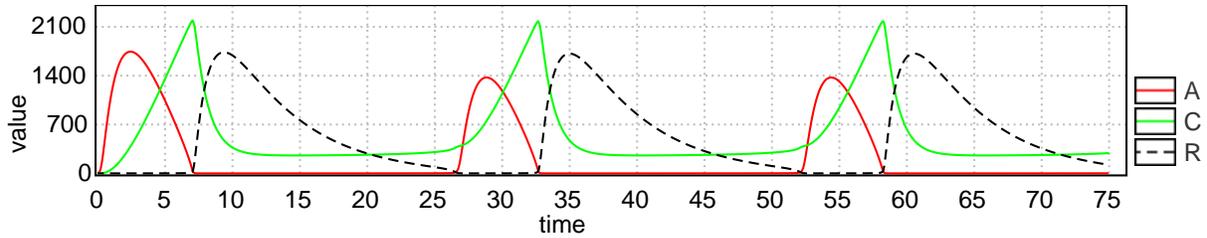}%
    \caption{ODE model simulation plot.}
    \label{fig:uppaal-ode-sim}
  \end{subfigure}
  \begin{subfigure}{\linewidth}
    \includegraphics[width=\linewidth]{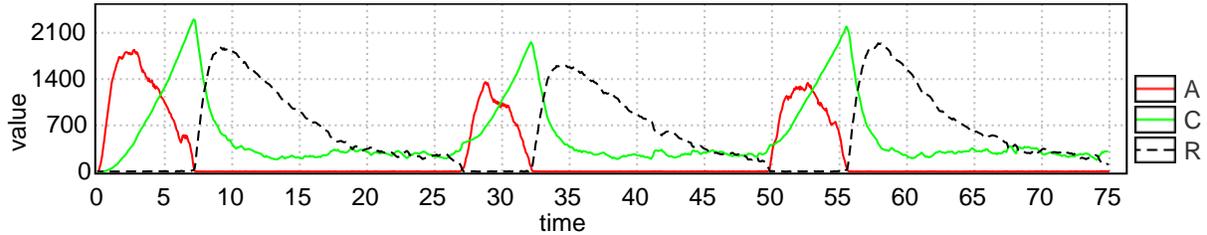}%
    \caption{Stochastic model simulation plot.}
    \label{fig:uppaal-sto-sim}
  \end{subfigure}
  \caption{\uppaalsmc simulations: {\tt simulate 1 [<=75] \{ A, C, R \}}.}
  \label{fig:uppaal-simulation}
\end{figure}

One of the key oscillatory behaviours in biology is the circadian rhythm that allows an organism to take advantage of periods of day and night to optimise when to maximise activity and recovery. 
We show how the genetic circadian oscillator of \cite{BarkaiLeibler2000,Vilar2002} can be modelled and analysed using \uppaalsmc. 
This synthetic model distils the essence of several real circadian oscillators and demonstrates how a reliable system can be constructed in the face of inherent stochasticity.
Figure~\ref{fig:ODE} shows a system of differential equations from~\cite{Vilar2002}. 
The equations are typeset in \uppaal as invariant expression on a location shown in Fig.~\ref{fig:uppaalode}, where each quantity $D_A, D_R, D'_A, M_A, M_R, A, R, C$ are modelled as continuous clock variables {\tt DA}, {\tt DR}, {\tt D\_A}, {\tt MA} etc with rates defined by a corresponding differential equation.
The preceding expressions about {\tt alphaA}, {\tt alpha\_A}, {\tt alphaR}, {\tt alpha\_R}, {\tt beta\_A} and so on are modelling the constants $\alpha_A, \alpha'_A, \alpha_R, \alpha'_R, \beta_A$ and so on.
The assignments on the first transition initialise all the variables with initial conditions.
\uppaalsmc is then used to simulate the model and provide plots of how the variable values evolve over time, which are displayed in Figure~\ref{fig:uppaal-ode-sim}.

\begin{figure}[!tb]
  \begin{subfigure}[b]{0.43\textwidth}
\begin{tabular}{l@{\hspace{5mm}} r}
\begin{minipage}[b]{0.4\linewidth}
{\scriptsize
\begin{eqnarray*}
\textnormal{A} + \textnormal{D}_A \stackrel{\gamma_{_A}}{\longrightarrow} \textnormal{D}_A'\label{eq:activationDA}\label{eq:firstreaction}\\
\textnormal{D}_A' \stackrel{\theta_{_A}}{\longrightarrow} \textnormal{D}_A + \textnormal{A}\\
\textnormal{A} + \textnormal{D}_R \stackrel{\alpha_{_R}}{\longrightarrow} \textnormal{D}_R'\label{eq:activationDR}\\
\textnormal{D}_R' \stackrel{\theta_{_R}}{\longrightarrow} \textnormal{D}_R + \textnormal{A}\\
\textnormal{D}_A' \stackrel{\alpha_{_{A}}'}{\longrightarrow} \textnormal{M}_A + \textnormal{D}_A'\label{eq:transcription_A}\\
\textnormal{D}_A \stackrel{\alpha_{_A}}{\longrightarrow} \textnormal{M}_A + \textnormal{D}_A\label{eq:transcriptionA}\\
\textnormal{D}_R' \stackrel{\alpha_{_{R}}'}{\longrightarrow} \textnormal{M}_R + \textnormal{D}_R'\label{eq:transcription_R}\\
\textnormal{D}_R \stackrel{\alpha_{_R}}{\longrightarrow} \textnormal{M}_R + \textnormal{D}_R\label{eq:transcriptionR}
\end{eqnarray*}}
\end{minipage}
& \begin{minipage}[b]{0.4\linewidth}
{\scriptsize
\begin{eqnarray*}
\textnormal{M}_A \stackrel{\beta_{_{_A}}}{\longrightarrow}\textnormal{M}_A + \textnormal{A}\label{eq:translationMA}\\
\textnormal{M}_R \stackrel{\beta_{_{_R}}}{\longrightarrow}\textnormal{M}_R + \textnormal{R}\label{eq:translationMR}\\
\textnormal{A} + \textnormal{R} \stackrel{\gamma_{_C}}{\longrightarrow}\textnormal{C}\label{eq:dimerisation}\\
\textnormal{C}\stackrel{\delta_{_A}}{\longrightarrow}\textnormal{R}\label{eq:decayC}\\
\textnormal{A} \stackrel{\delta_{_A}}{\longrightarrow}\emptyset\label{eq:decayA}\\
\textnormal{R} \stackrel{\delta_{_R}}{\longrightarrow}\emptyset\label{eq:decayR}\\
\textnormal{M}_A \stackrel{\delta_{_{M_A}}}{\longrightarrow}\emptyset\label{eq:decayMA}\\
\textnormal{M}_R \stackrel{\delta_{_{M_R}}}{\longrightarrow}\emptyset\label{eq:decayMR}\label{eq:lastreaction}
\end{eqnarray*}}
\end{minipage}
\end{tabular}
\vspace*{-2mm}
\caption{Reactions.}
\label{fig:reactions}
\end{subfigure}
\begin{subfigure}[b]{0.5\textwidth}
\begin{tabular}{ l l l l}
    \includegraphics[height=10ex]{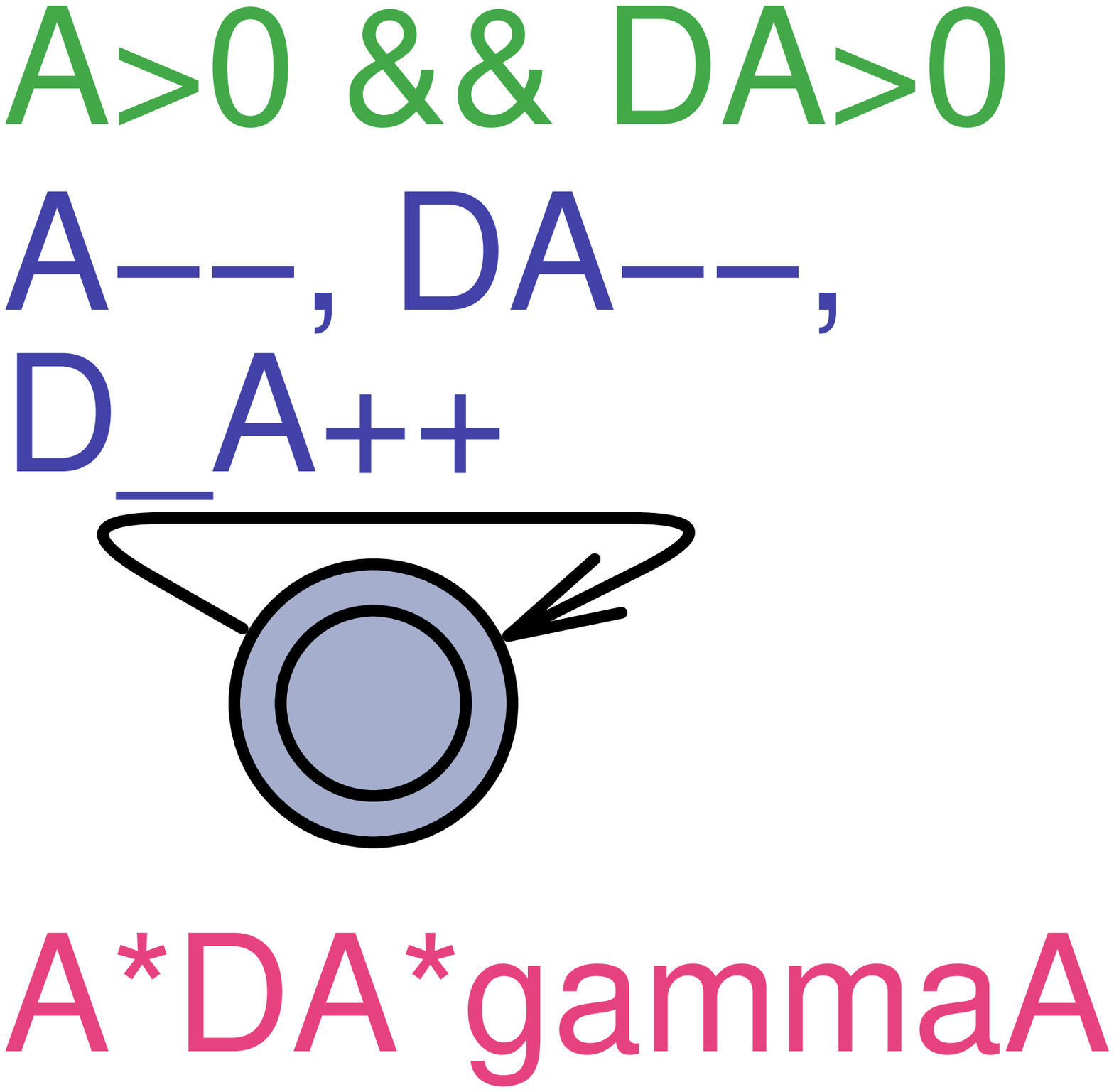} &
    \includegraphics[height=10ex]{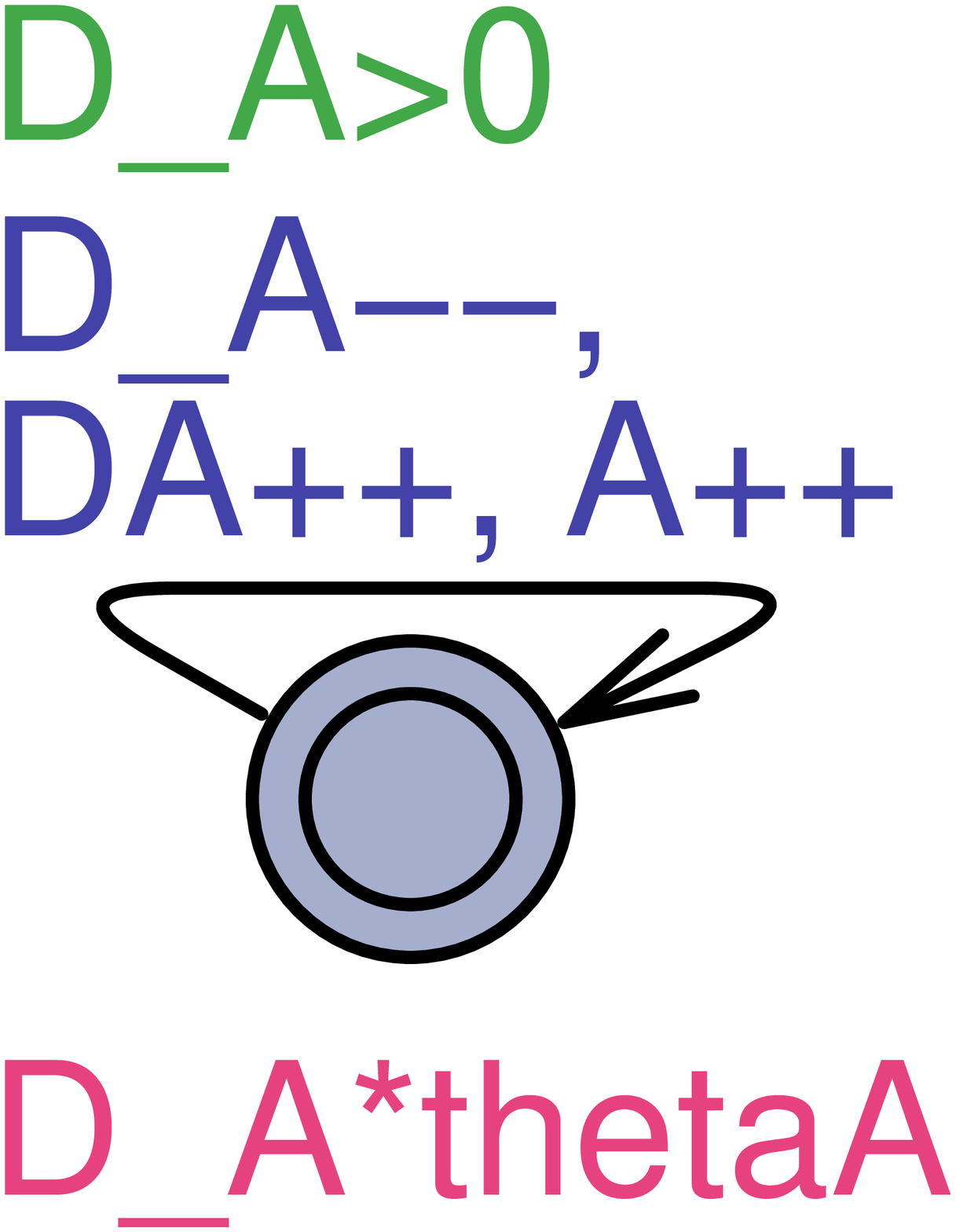} &
    \includegraphics[height=10ex]{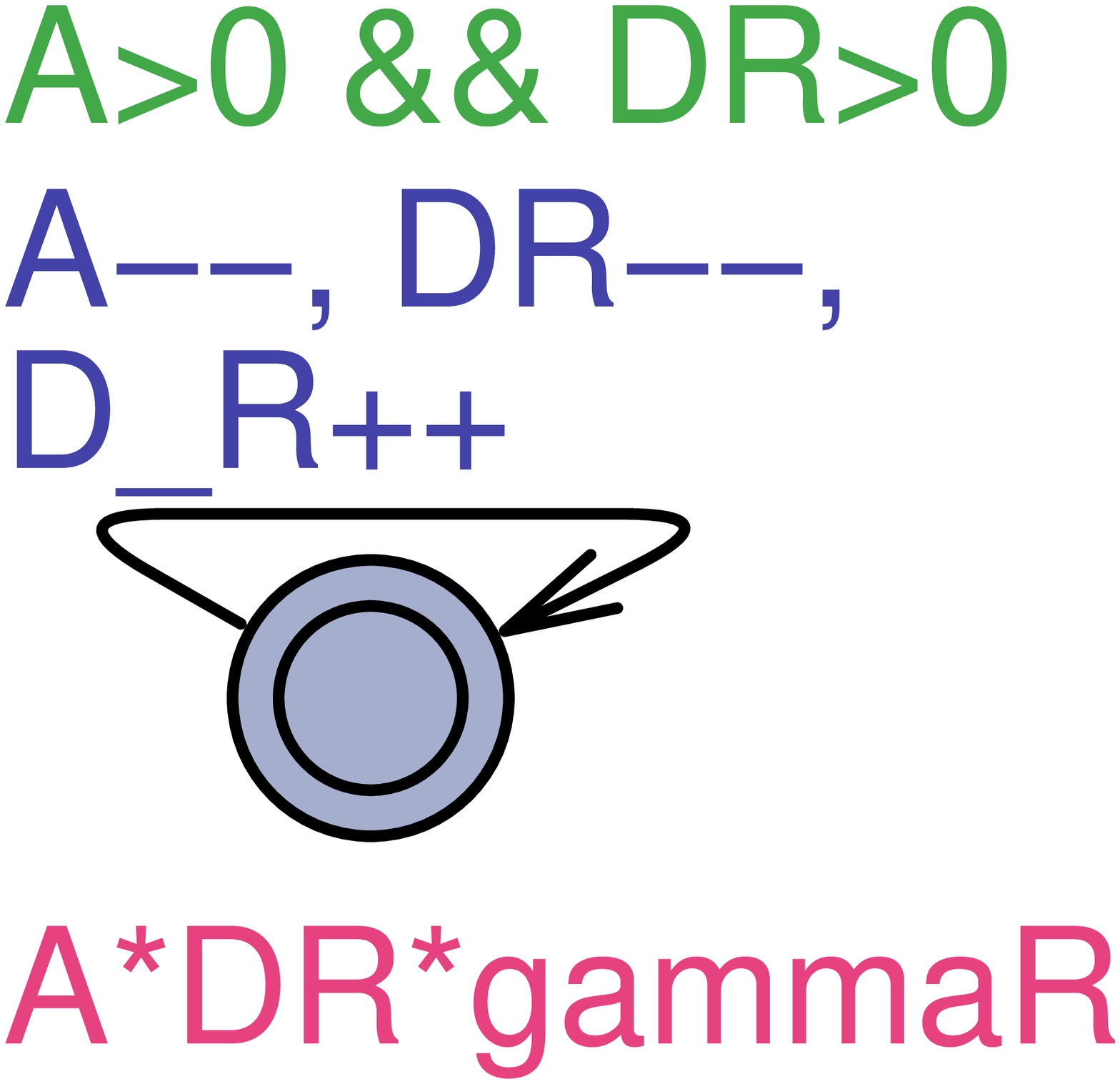} &
    \includegraphics[height=10ex]{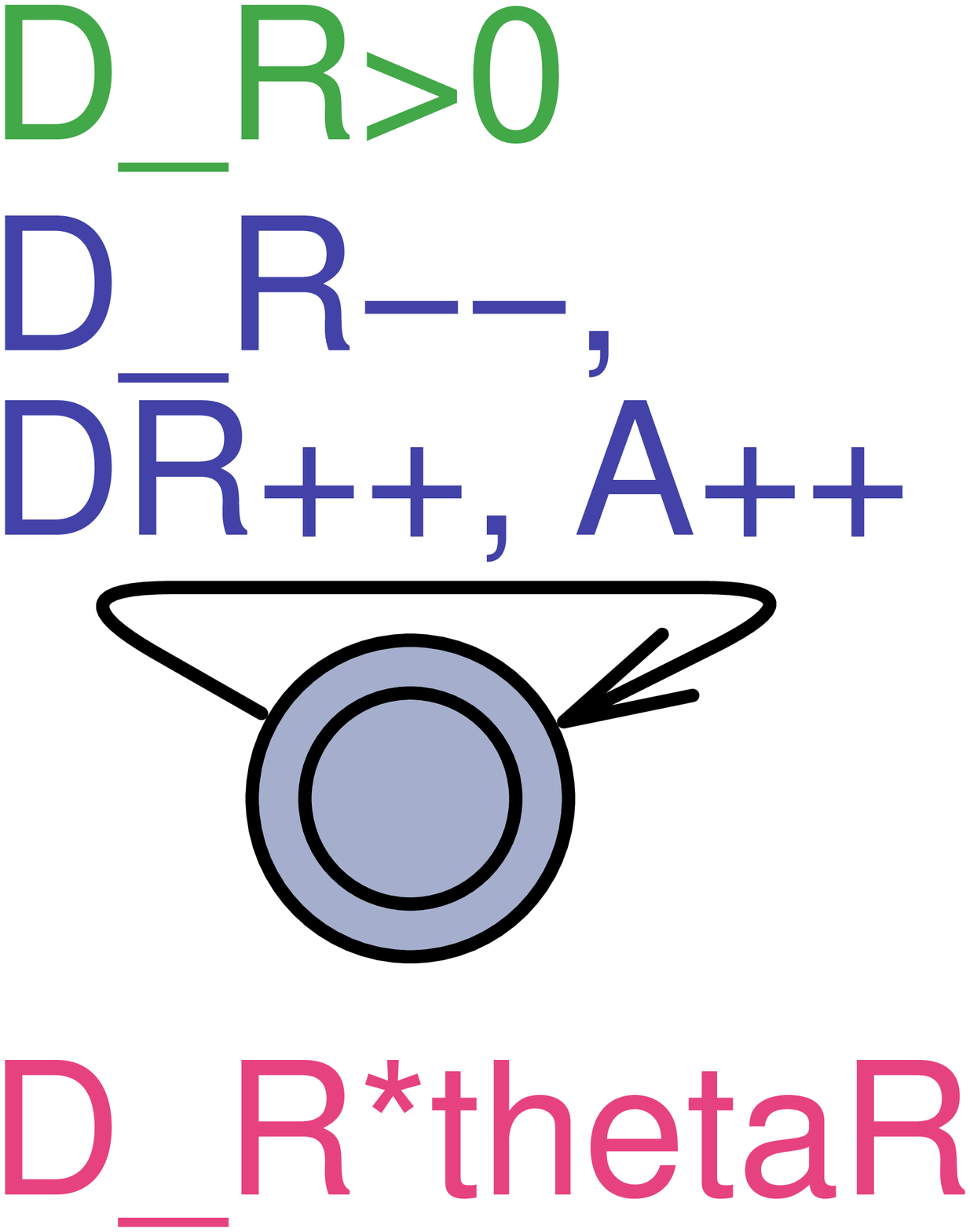} \\
  \includegraphics[height=8.5ex]{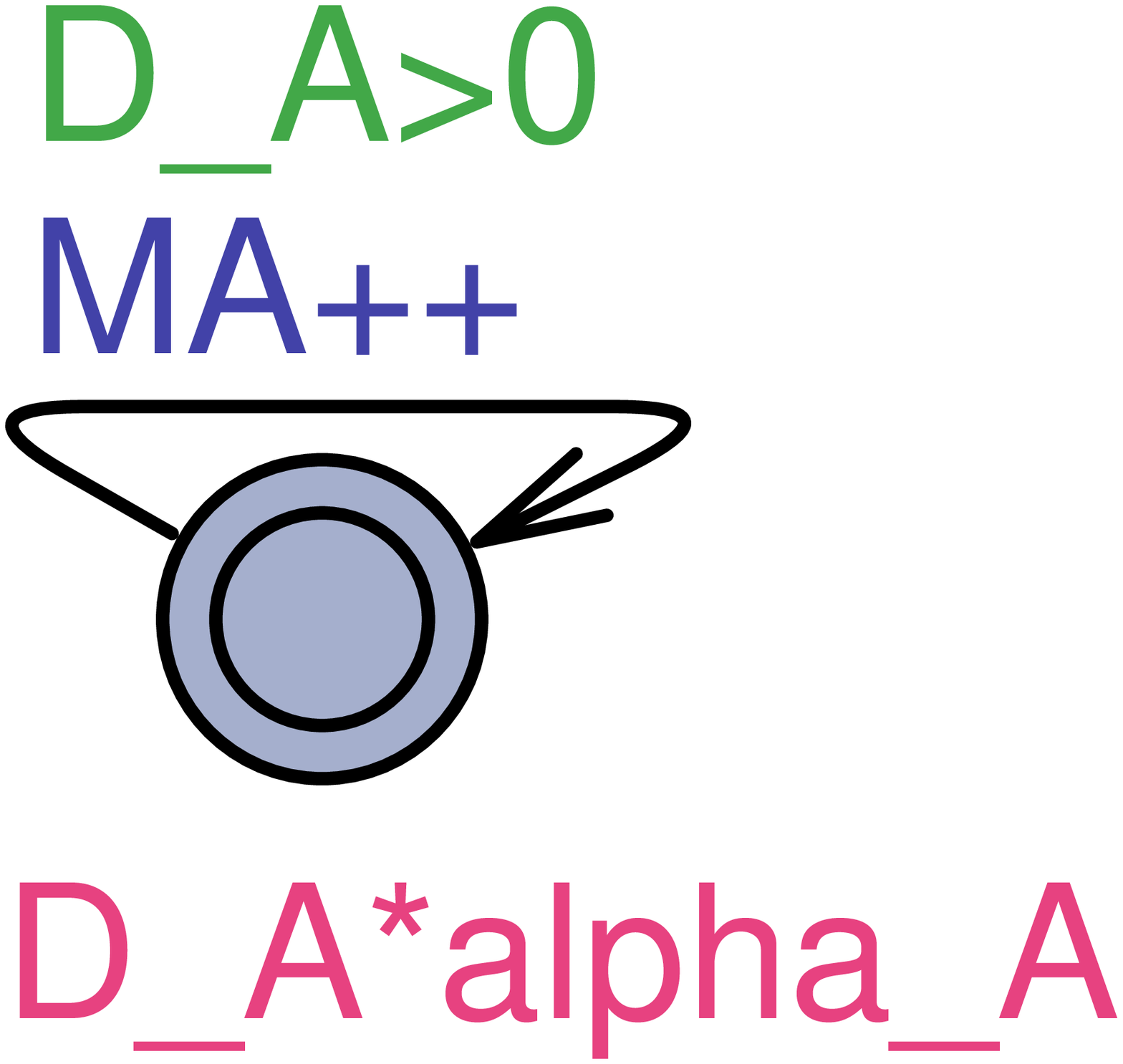} &
    \includegraphics[height=8.5ex]{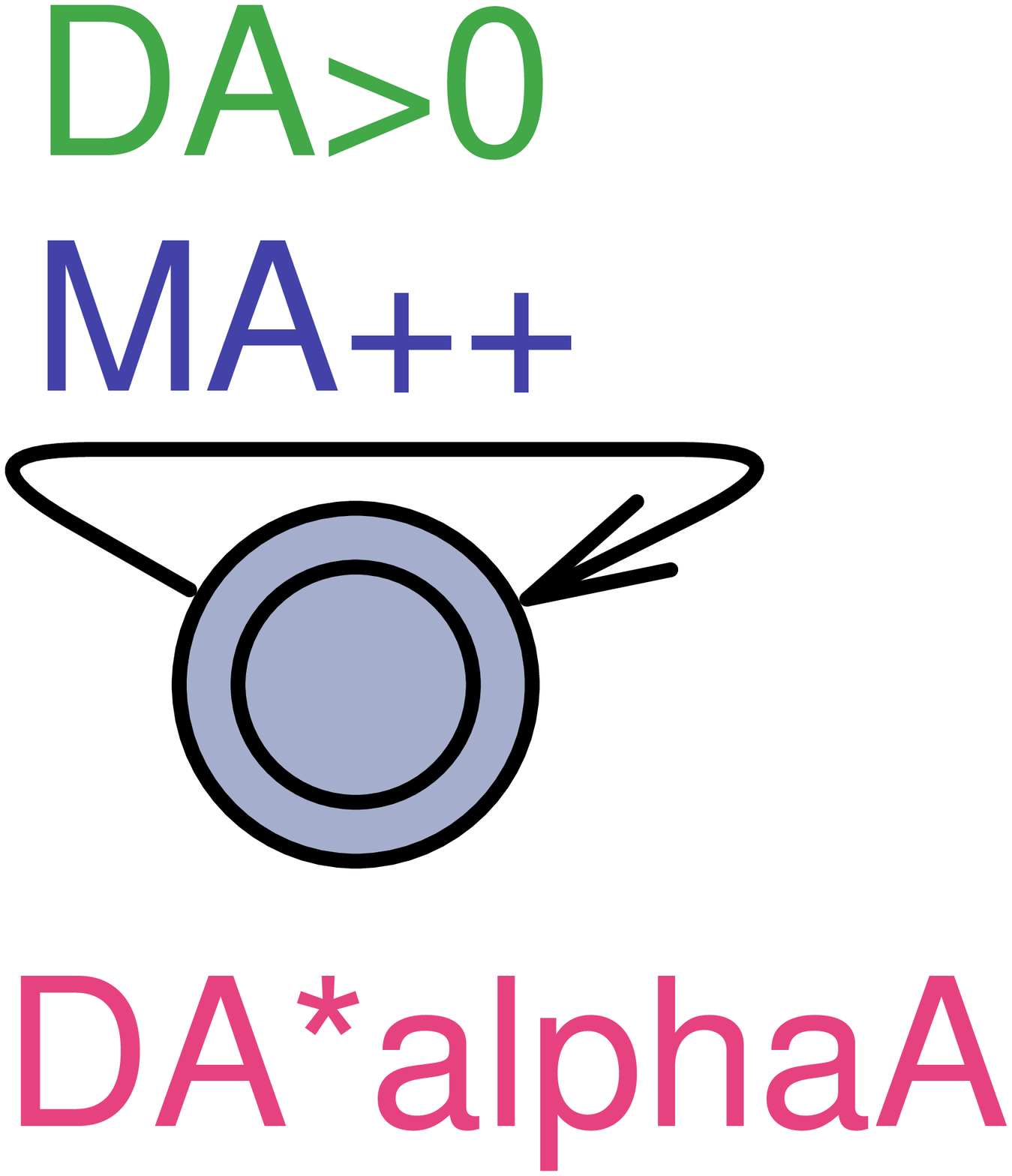} &
   \includegraphics[height=8.5ex]{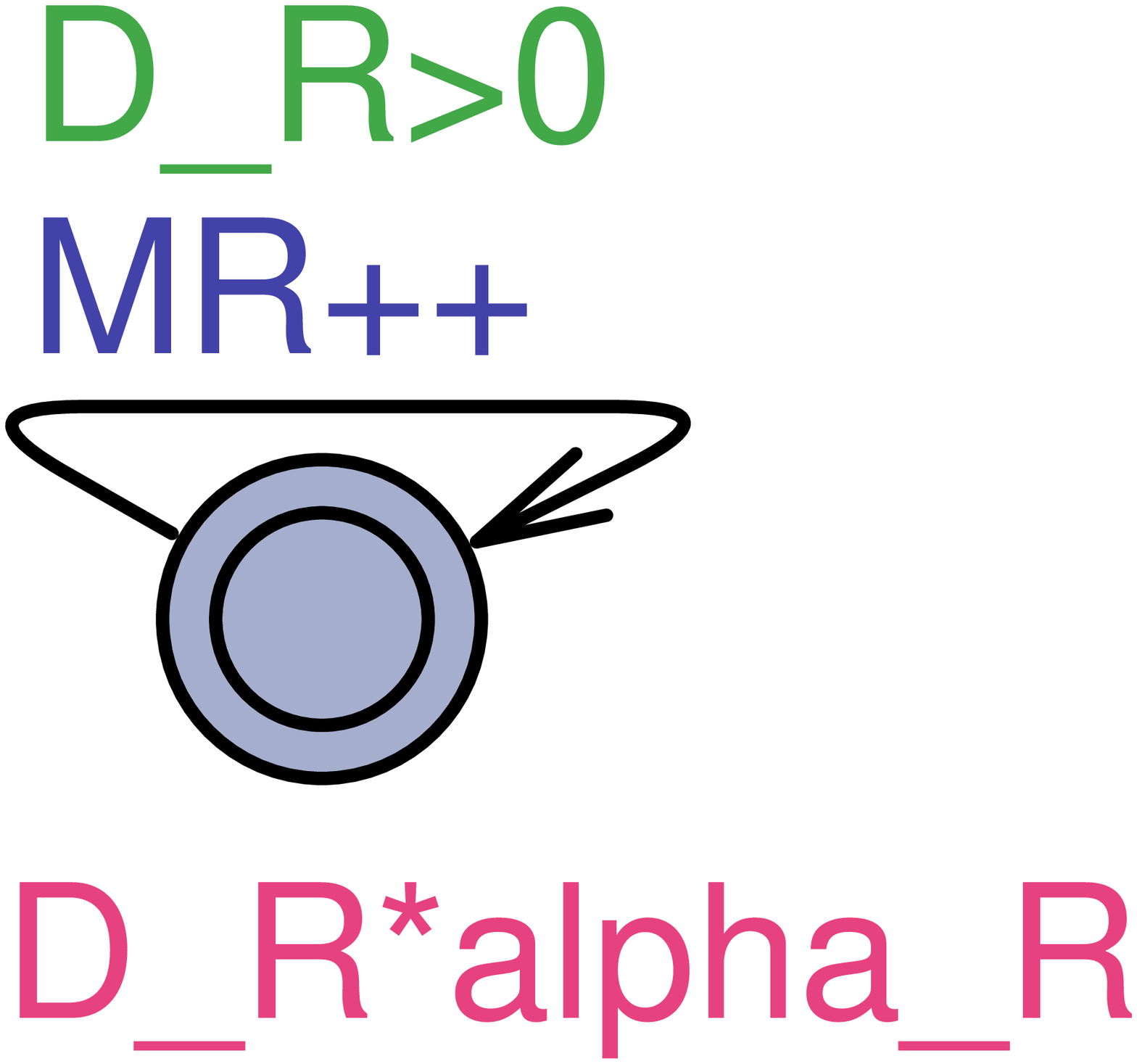} &
  \includegraphics[height=8.5ex]{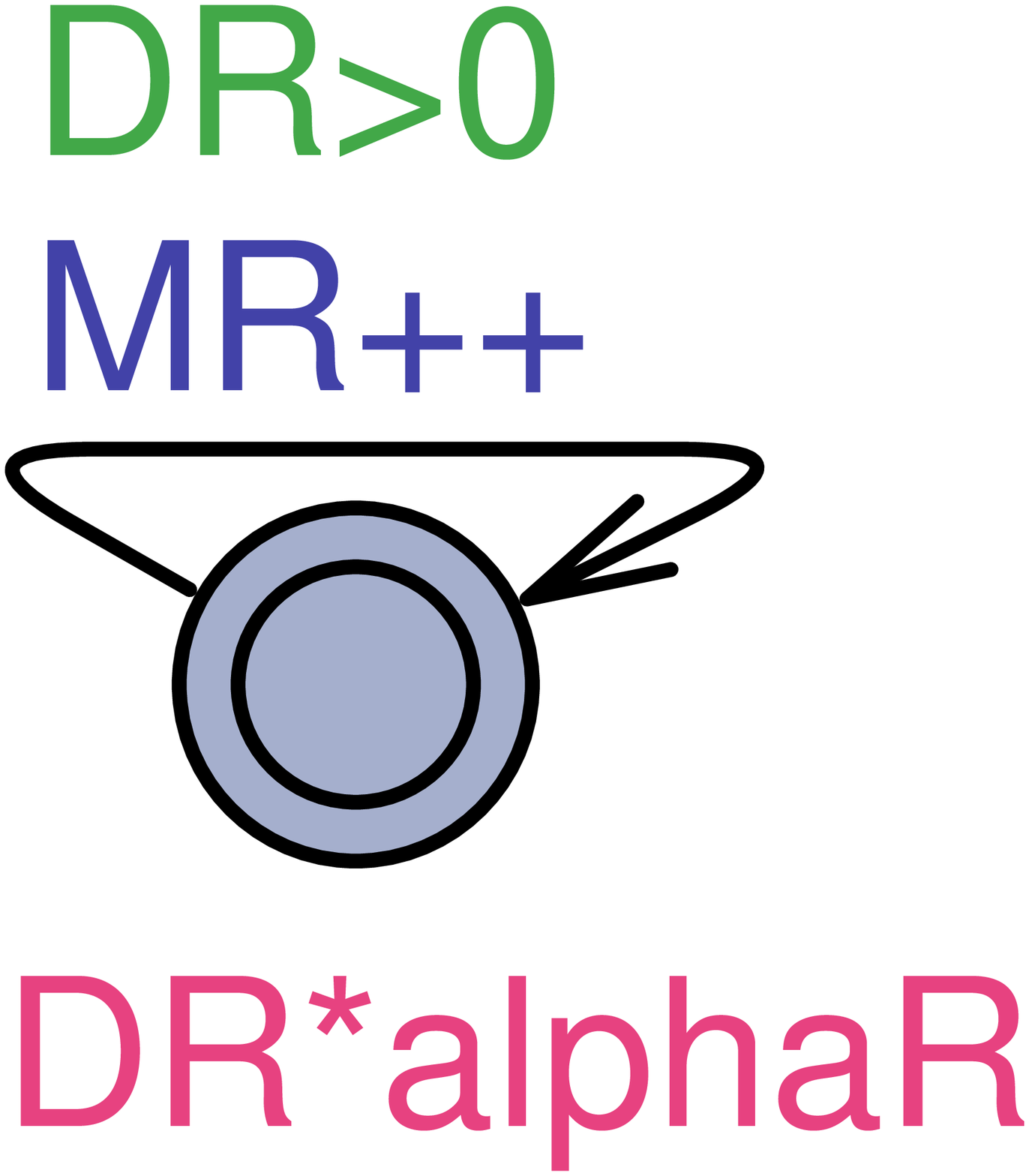} \\
    \includegraphics[height=8.5ex]{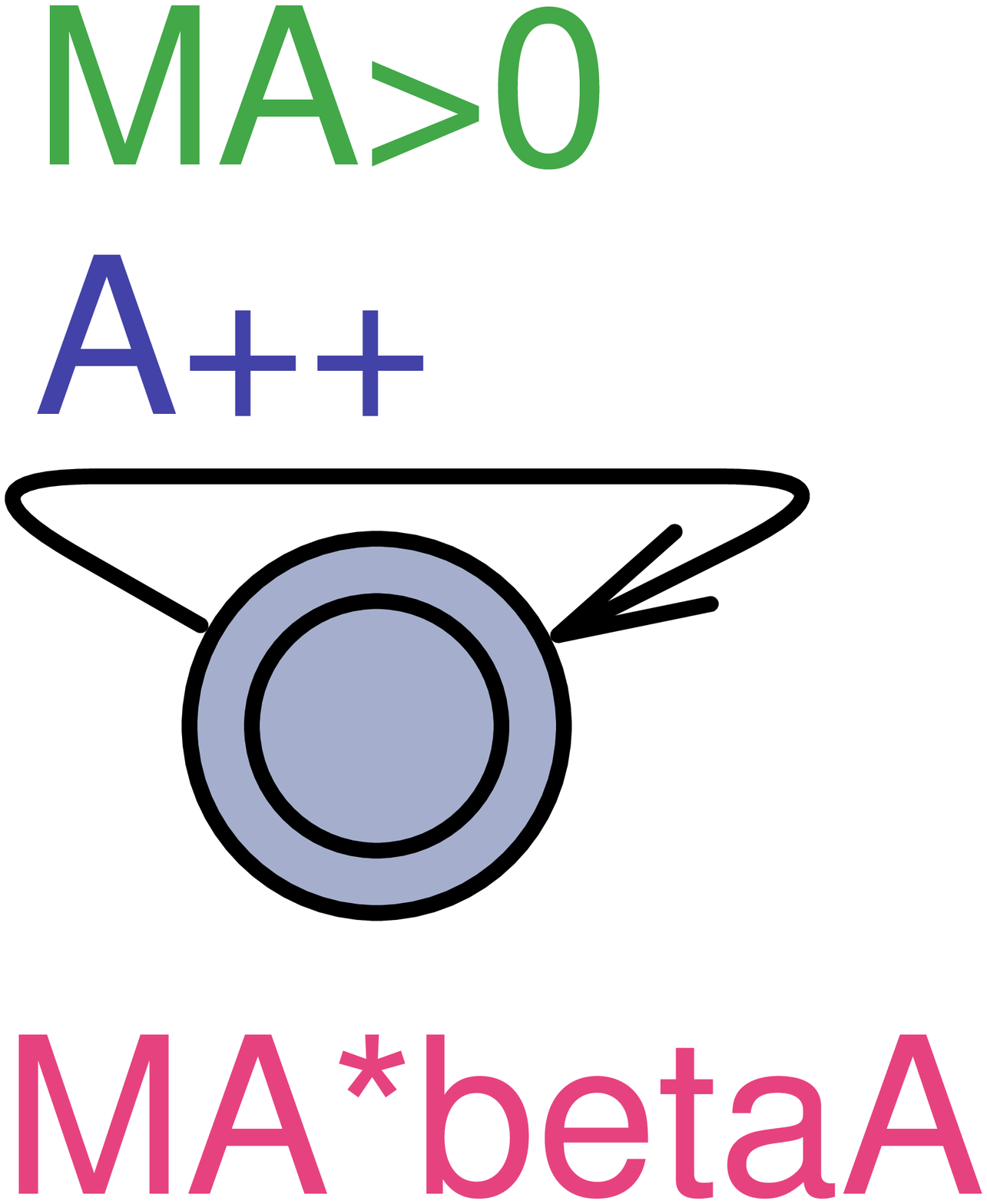} &
    \includegraphics[height=8.5ex]{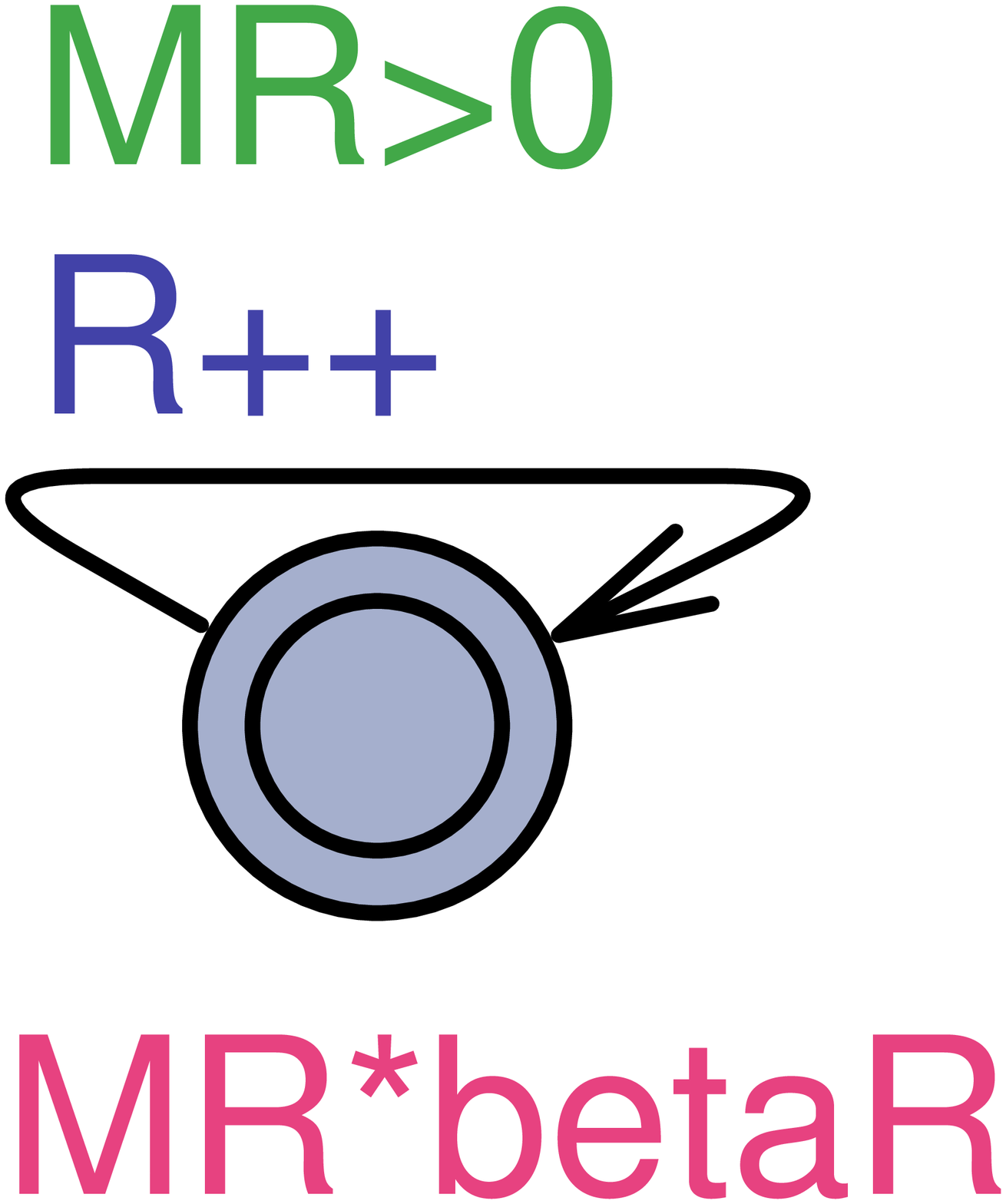} &
    \includegraphics[height=10ex]{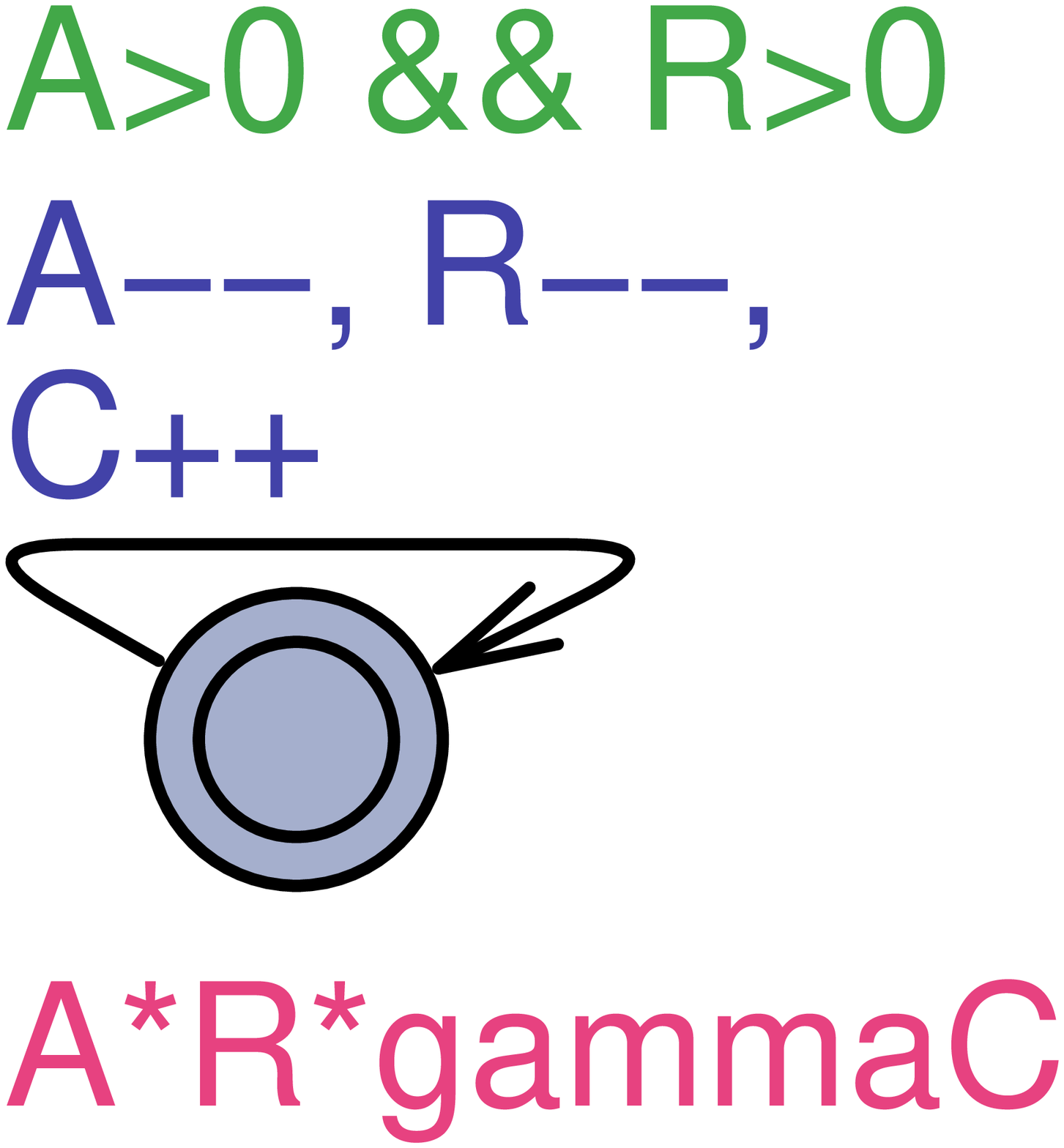} &
    \includegraphics[height=8.5ex]{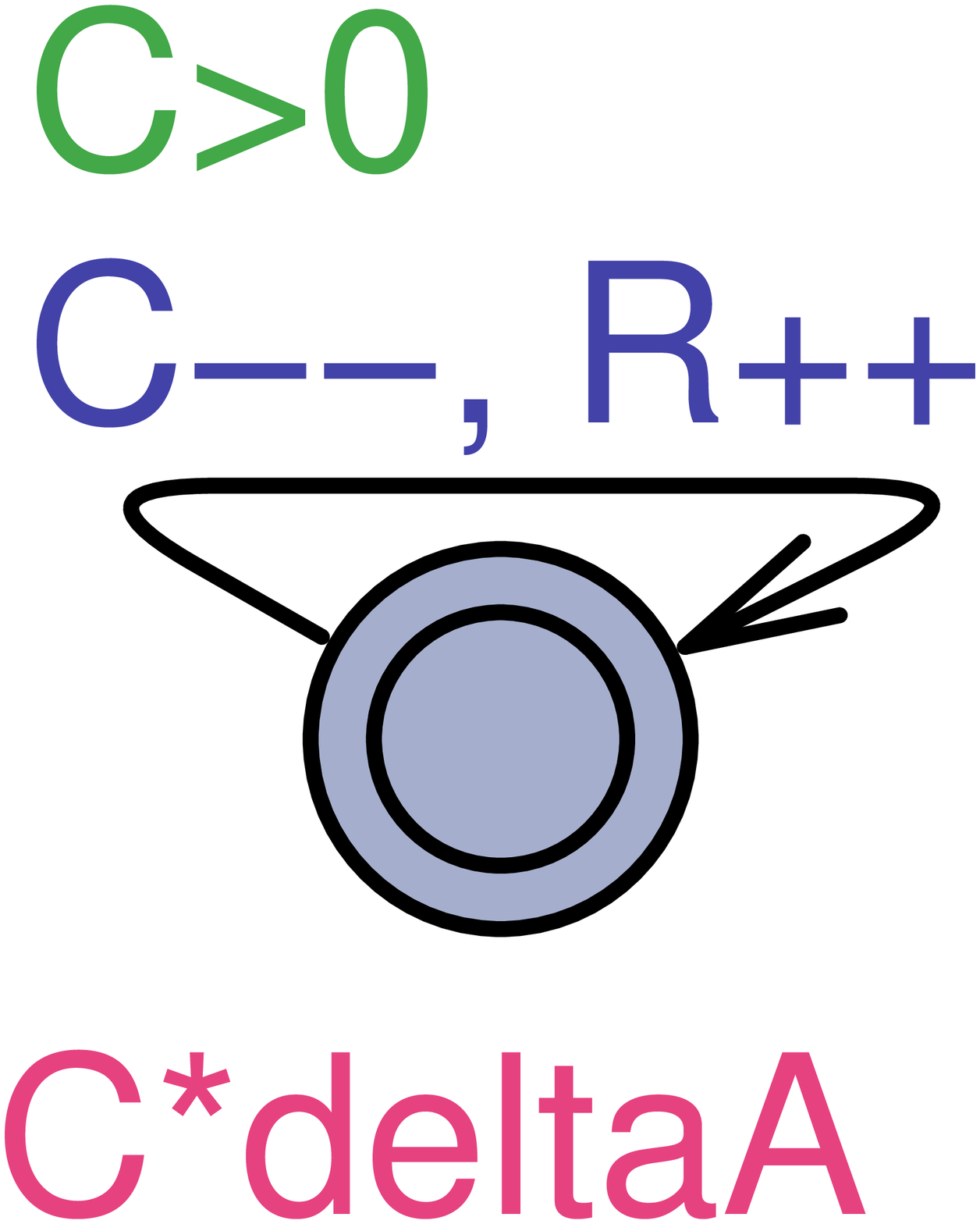} \\
    \includegraphics[height=8.5ex]{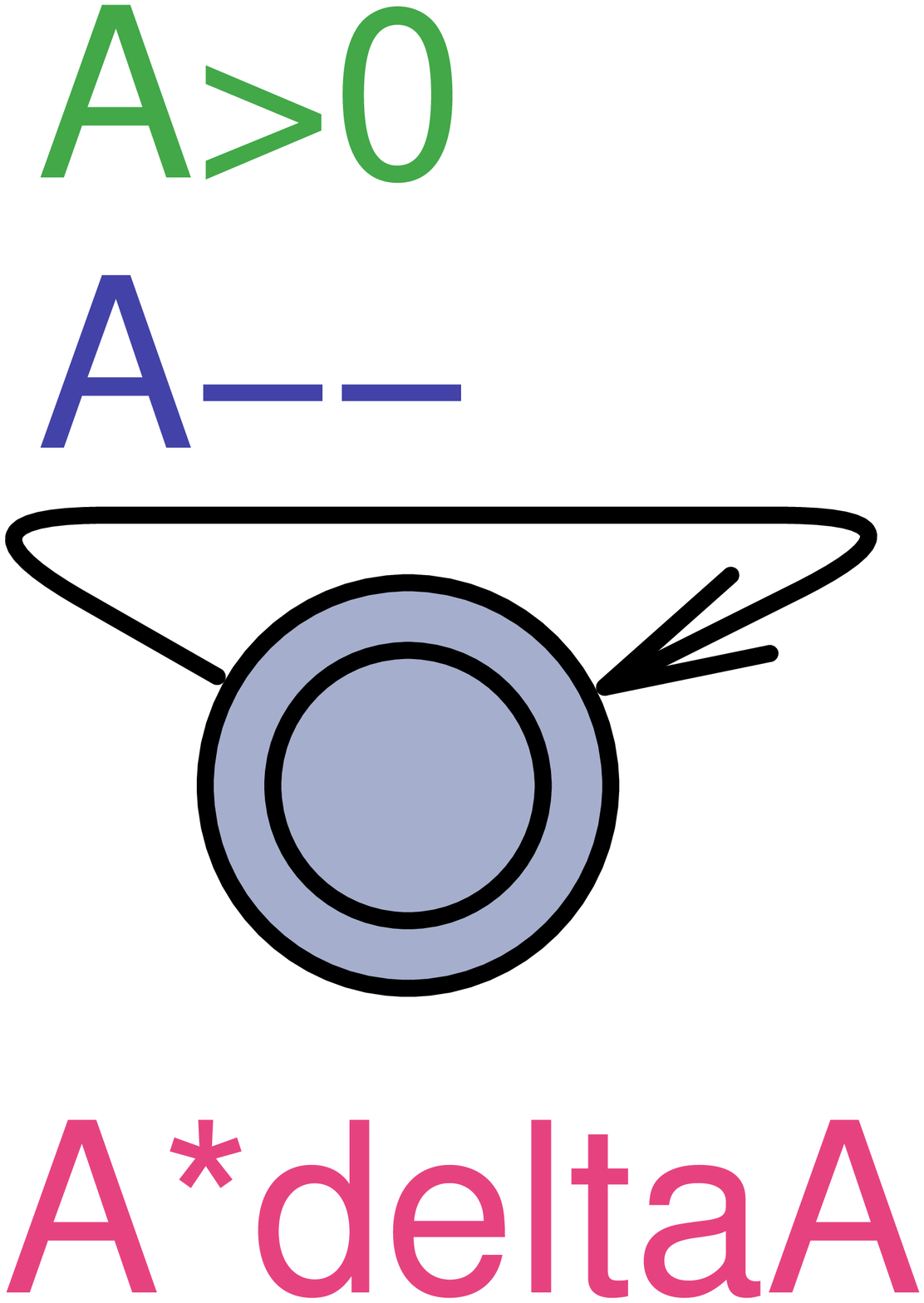} &
    \includegraphics[height=8.5ex]{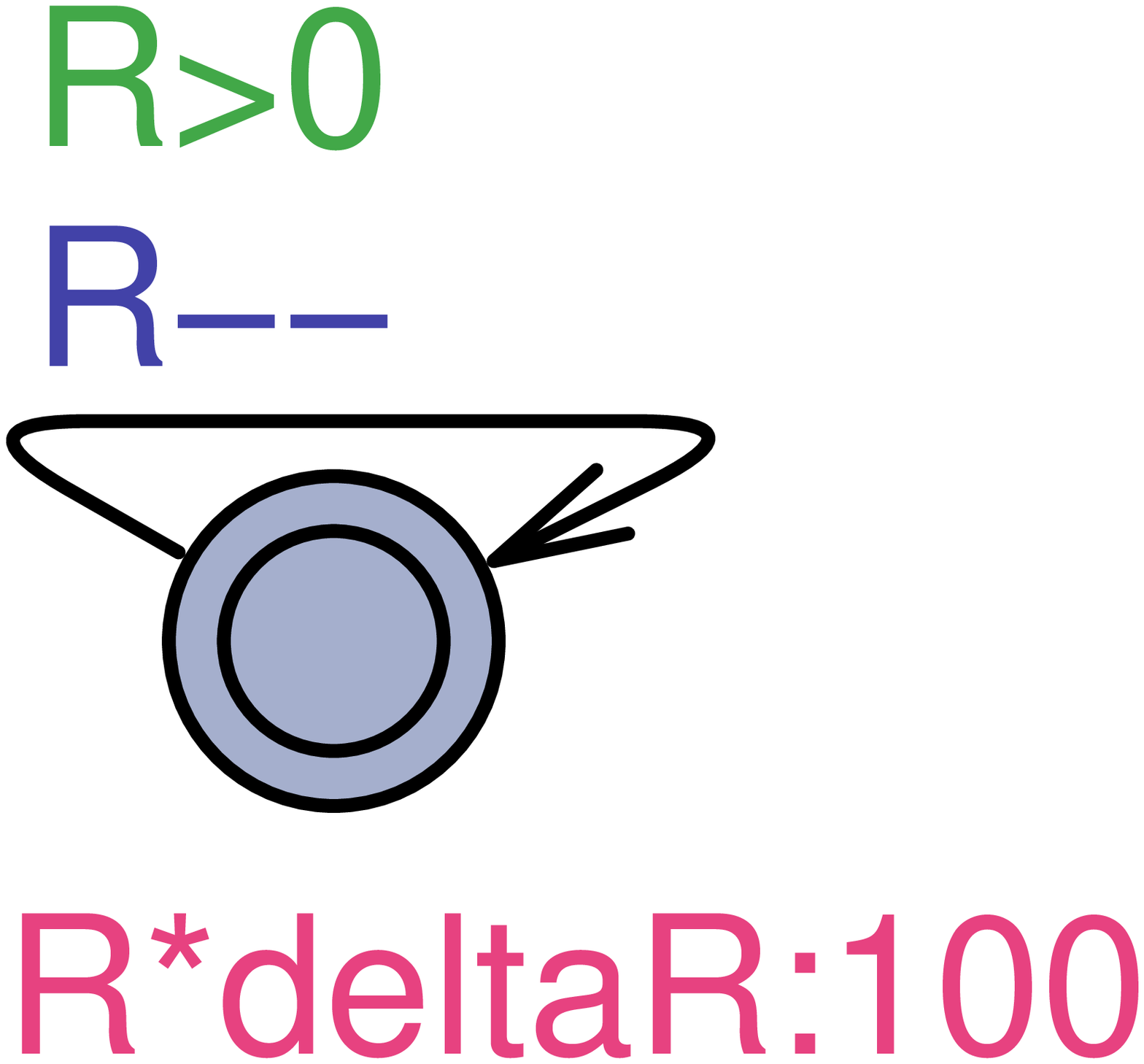} &
    \includegraphics[height=8.5ex]{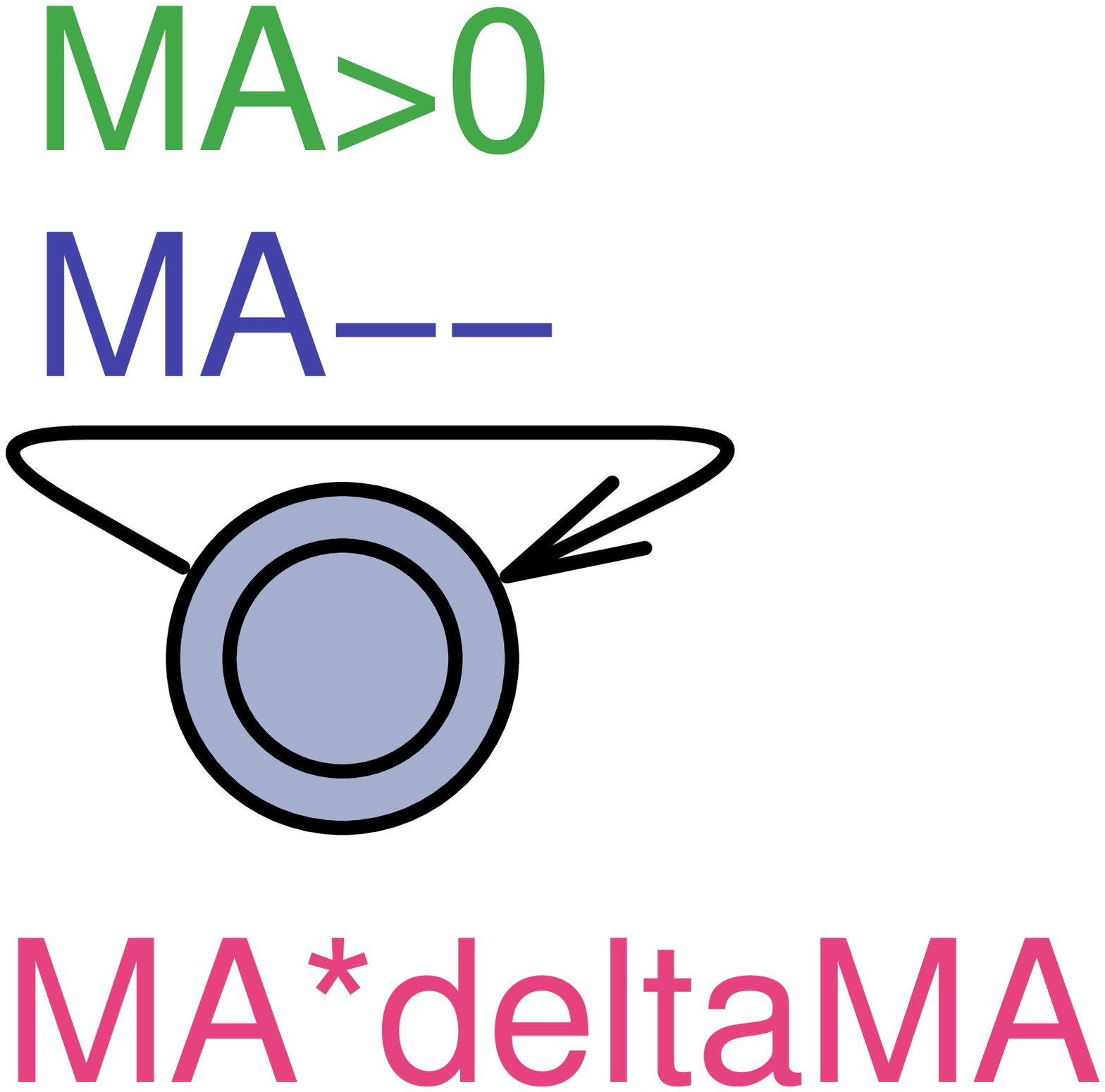} &
    \includegraphics[height=8.5ex]{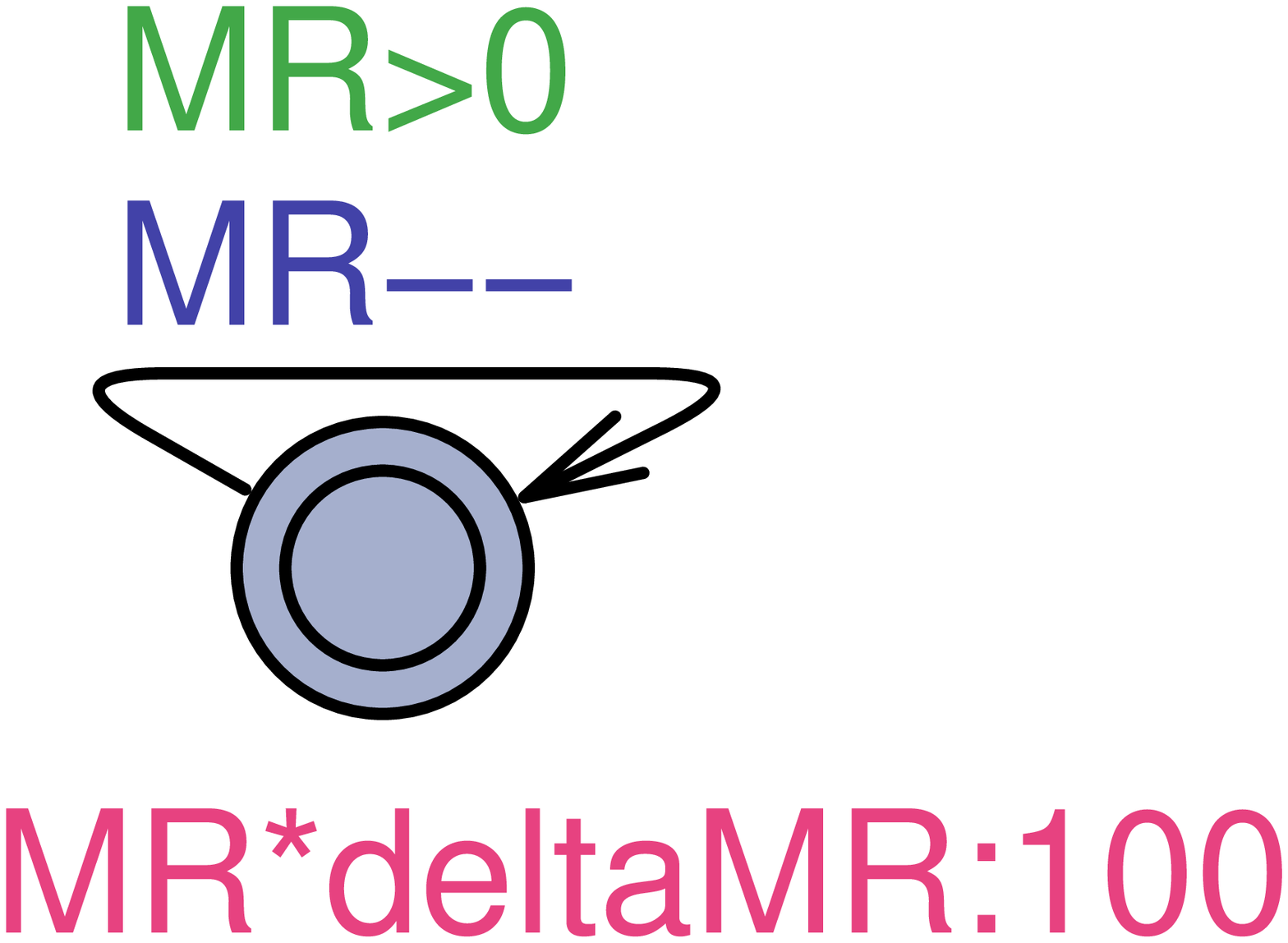}
\end{tabular}
    \caption{\uppaal model representation.}
    \label{fig:uppaalstochastic}
\end{subfigure}
\caption{Stochastic model of genetic oscilator.}
\end{figure}

The ODE system can be interpreted as a system behavior near the thermodynamic limit (infinite population sizes while maintaining the same concentrations).
Alternatively this oscilator can be modeled as a system of stochastic chemical reactions, where each molecule is counted as discrete entity moving according to Brownian motion laws.
Using a standard translation between deterministic and stochastic semantics of chemically reacting systems (e.g., Gillespie's algorithm \cite{Gillespie1977}) the coefficients in ODE can be interpreted as reaction rates.
The reactions are enumerated in Fig.~\ref{fig:reactions}.
Each reaction is then modeled as a separate SHA process shown in Fig.~\ref{fig:uppaalstochastic}, which can be viewed as an encoding of continuous-time Markov chain process.
For example, the first reaction means taking one molecule of each $A$ and $D_A$ substance and producing $D'_A$ with a rate of $\gamma_A$, which can be interpreted as a transition requiring positive amount of {\tt A} and {\tt DA} (modeled as integer variables), consuming one molecule each and producing one {\tt D\_A} and the reaction rate is {\tt gammaA} and proportional to available quantities of {\tt A} and {\tt DA}.
The resulting trajectories of quantities {\tt A}, {\tt C} and {\tt R} are displayed in Fig.~\ref{fig:uppaal-sto-sim}, where the patterns seem to resemble the one from ODE model (Fig.~\ref{fig:uppaal-ode-sim}), but the inherent stochasticity result in shaky lines (even saw teeth) and unpredictably fluctuating amplitudes and their periods.
Interestingly, our formalism is flexible enough to accommodate a hybrid model combining stochastic aspects as in Fig.~\ref{fig:uppaal-sto-sim} \emph{and} continuous aspects as in Fig.~\ref{fig:uppaal-ode-sim}.

\begin{figure}[!htb]
  \begin{tabular}{ccc}
    \includegraphics[width=0.31\linewidth]{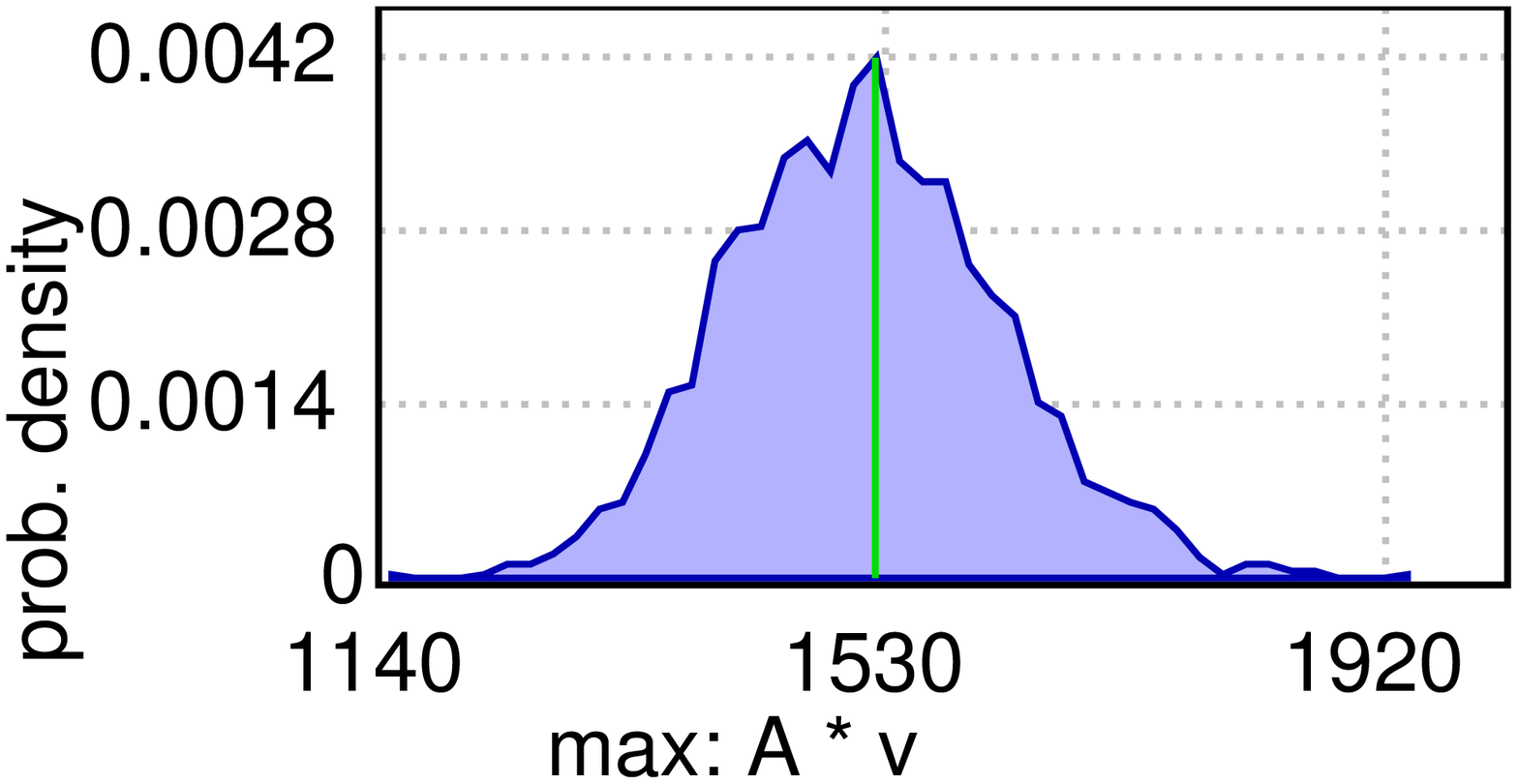}&
    \includegraphics[width=0.31\linewidth]{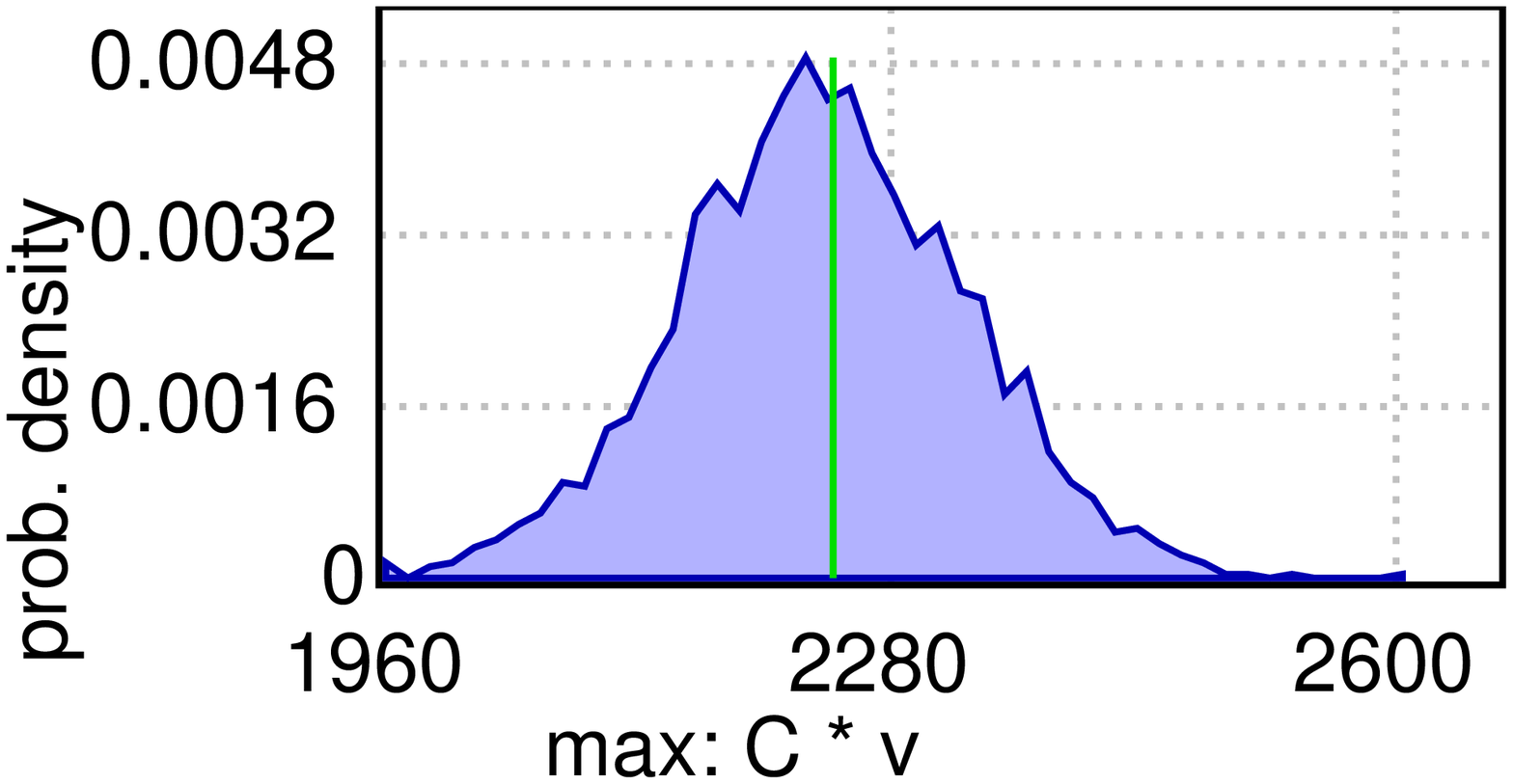} &
    \includegraphics[width=0.31\linewidth]{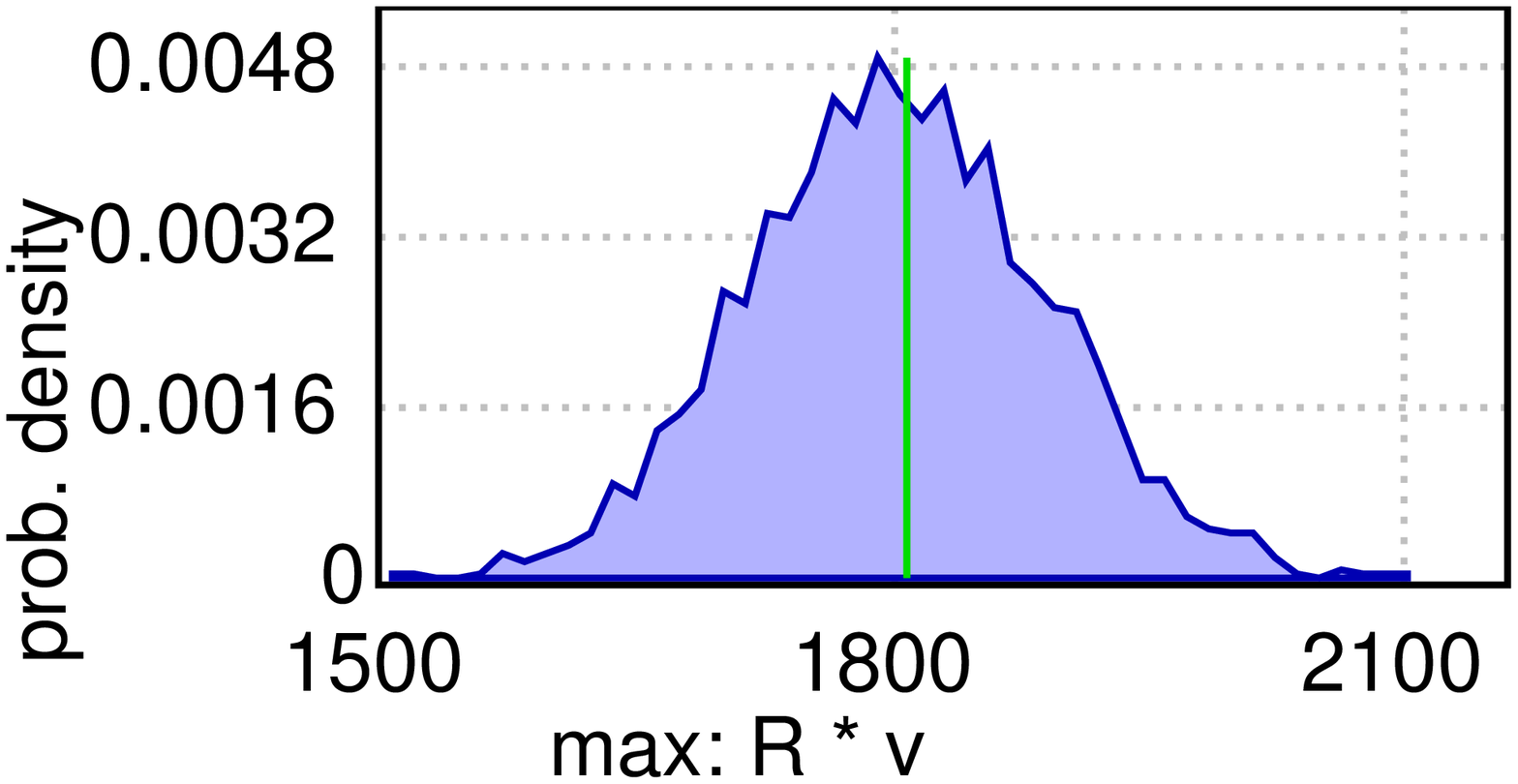} \\
    \includegraphics[width=0.31\linewidth]{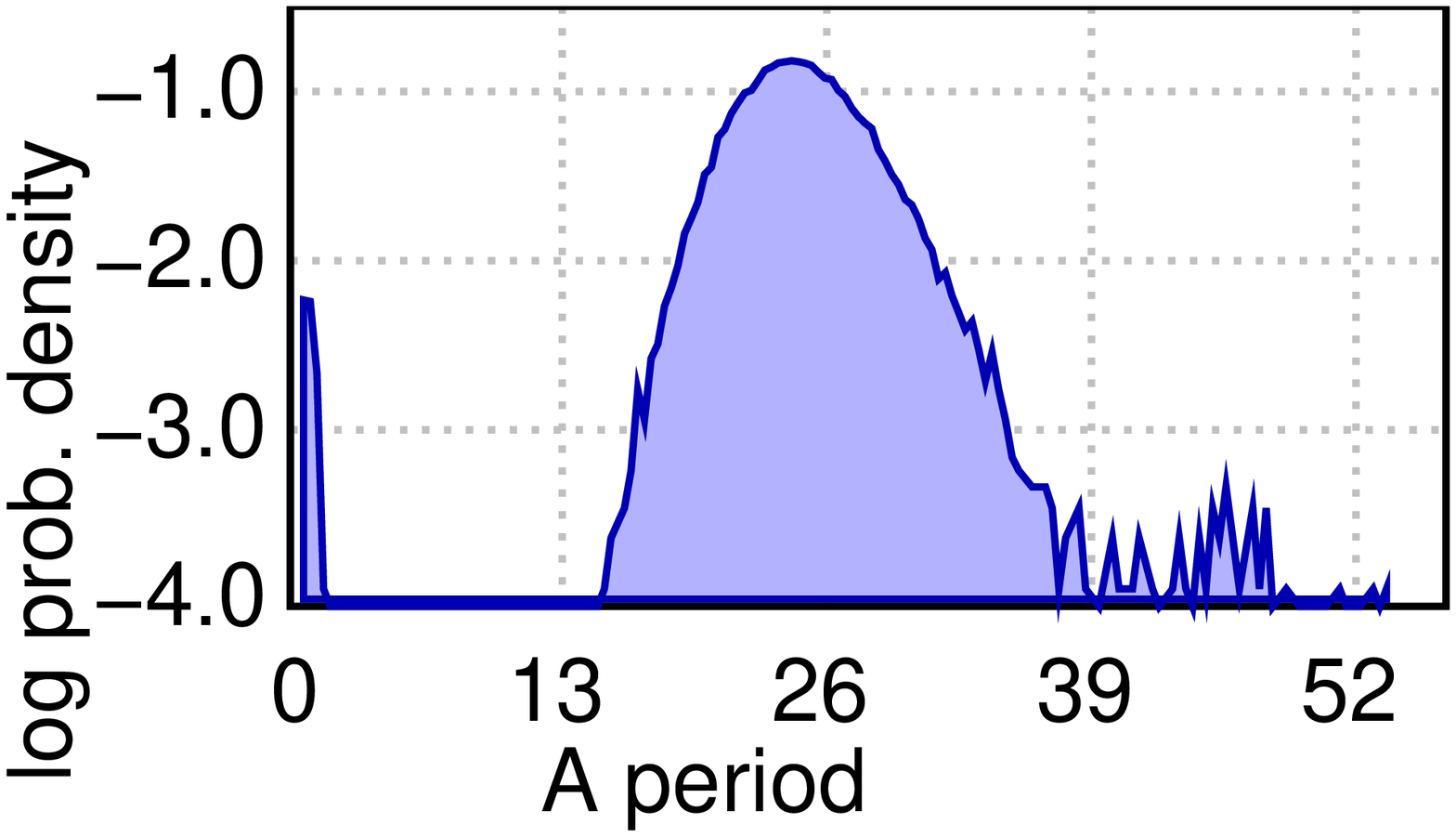} &
    \includegraphics[width=0.31\linewidth]{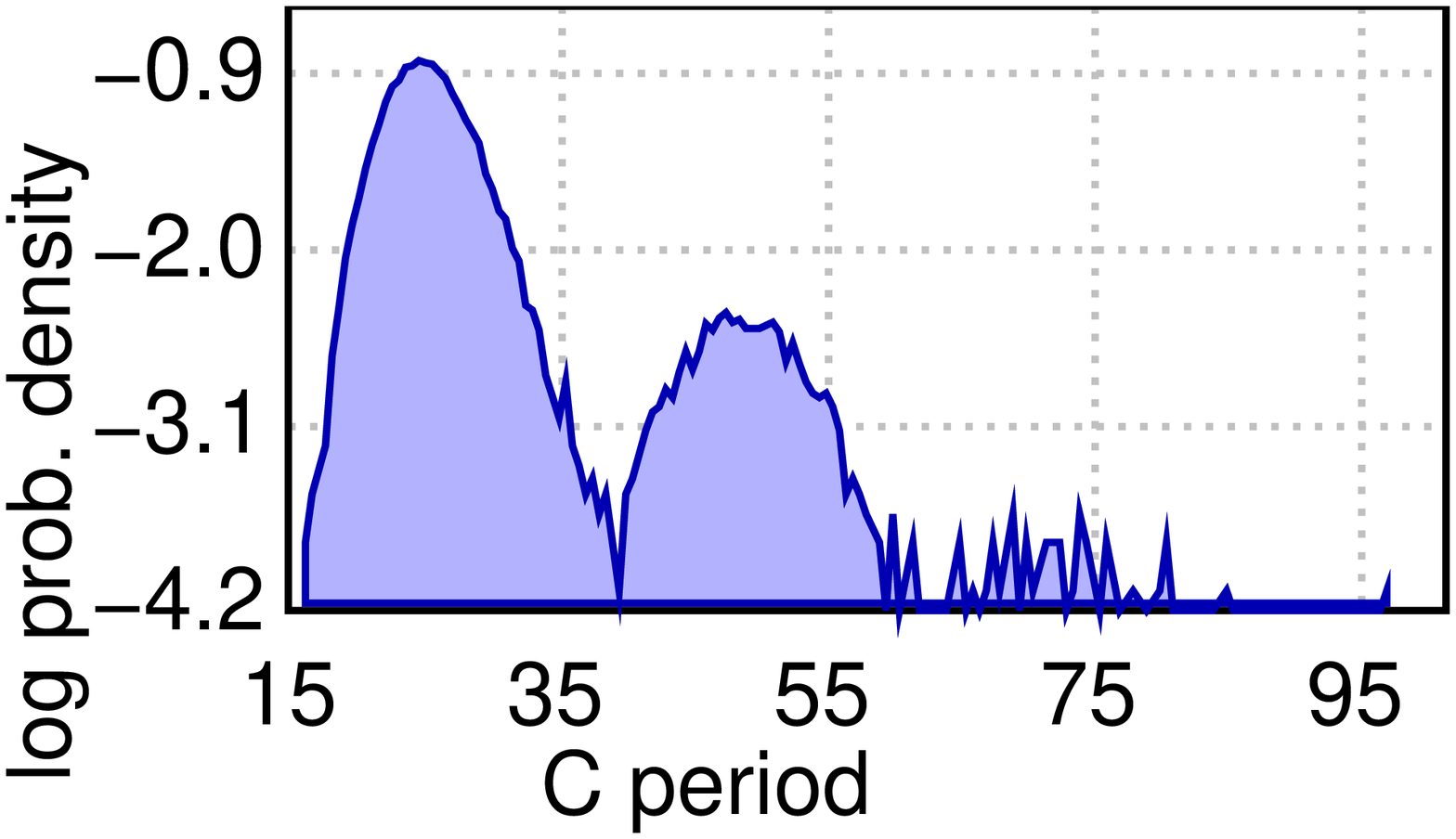} &
    \includegraphics[width=0.31\linewidth]{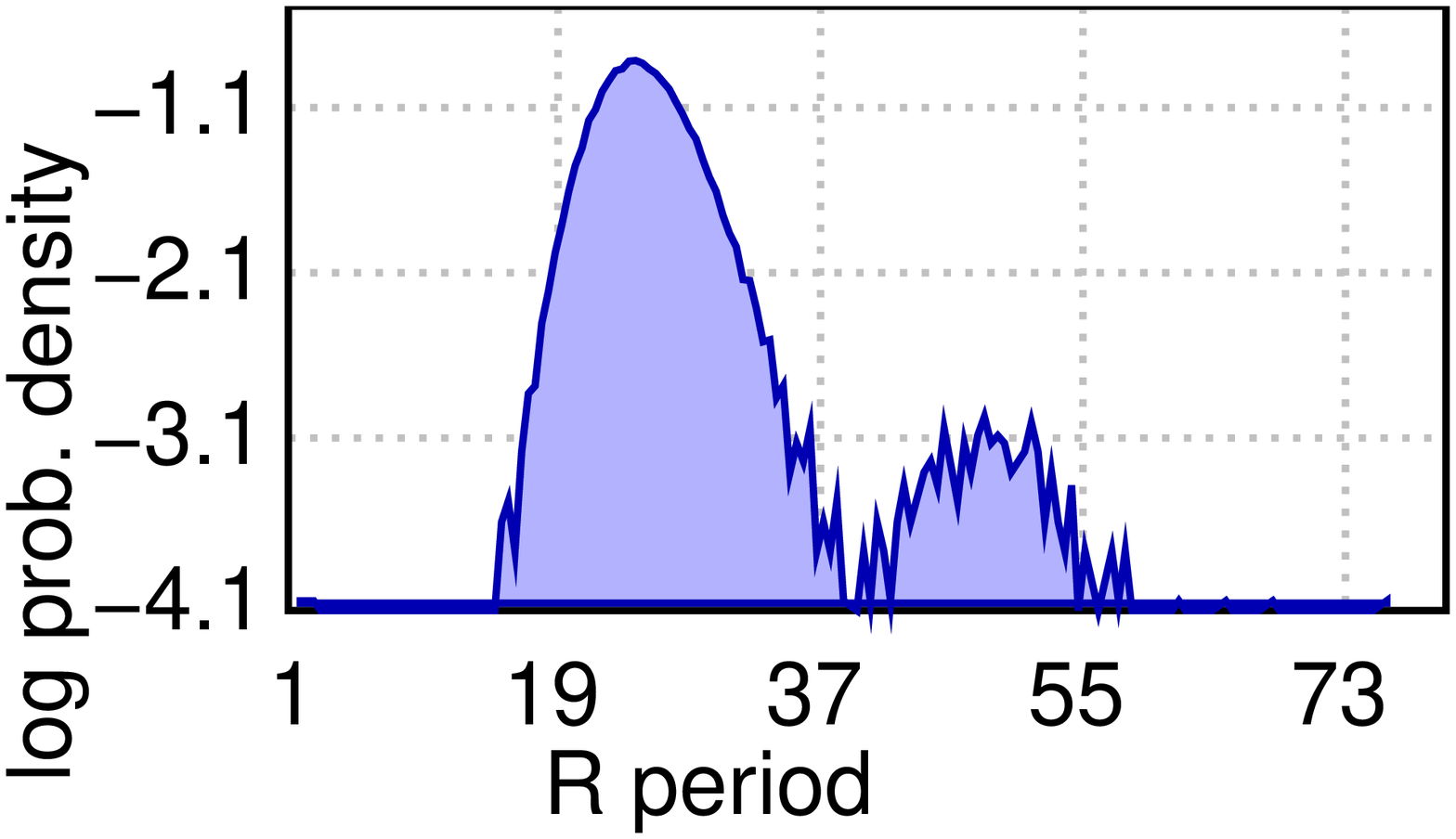} \\
    (a) {\tt A}. & (b) {\tt C}. & (c) {\tt R}.
  \end{tabular}
  \caption{Estimated probability density distributions for amplitude and period.}
  \label{fig:ampper}
\end{figure}

The amplitude of each protein quantity can be measured by the queries
{\tt E[<=75; 2000](max: $q$)}, where 75 is the time limit for simulation, 2000 is the number of simulations and $q$ is either {\tt A}, {\tt C}, or {\tt R}.
The upper plots of Fig.~\ref{fig:ampper} show the probability density for a range of possible values of amplitude with a vertical line for average value.

\uppaalsmc  can  also  estimate  a  distance between  peaks  by  using
techniques  developed for MITL  (Metric Interval  Temporal Logic  -- a
more expressive property language than a subset of CTL supported by \uppaal).  
The idea of  the approach is to  translate MITL formula  into a monitoring
automata which start an auxiliary clock  {\tt x} with a first peak and
stop  with a second  peak \cite{MITLSMC}. 
To detect peaks of {\tt A} when its amount rises above 1100 and drops below 1000
within 5 time units, we use the formula (in the tool syntax):
{\tt true U [<=1000] (A>1100 \& true U[<=5] A<=1000)}.
Then the  distance between
peaks can  be estimated by measuring  maximal value of  clock {\tt x}.
The result  is shown  as logarithm of  probability density plots  in a
second  row of  Fig.~\ref{fig:ampper}.  The  plots show  that  in most
cases  the  measured  distance  between  peaks  is  about  24.2  hours
(slightly more than one day-night cycle).  Then there are some smaller
bumps with several magnitudes lower probability which can be explained
by either  a) false  positive peak  as MITL monitor  is confused  by a
sudden stochastic saw tooth in signal {\tt A}, or b) missing a peak or
two, or even  three (in {\tt C}) if  the peak is not high  enough to be
registered, hence the next one is registered instead.


\section{Conclusions}
\label{con-sec}

This  paper presents  an extensions  of the  \uppaalsmc  model checker
supporting statistical model  checking for stochastic hybrid automata,
where dynamic of systems can be  specified with ODEs.  This is a major
advance in  comparison to existing  SMC checkers that can  only handle
simple derivatives. Currently, the  tool applies the Euler integration
method. In the future, we  plan to implement more robust methods, such
as Runge-Kutta's. Another contribution  will be to support rare-events
as it is the case in \cite{ZBC12,Sedwards2012,Jegourel2012}.

\bibliographystyle{eptcs}
\bibliography{bibliography}
\end{document}